\documentclass[11pt]{revtex4-1}
\usepackage{graphicx,fullpage}
\usepackage{epsfig}
\usepackage{xcolor}
\usepackage{amsmath}
\usepackage{amssymb}
\usepackage{bm}
\usepackage{array}
\usepackage{booktabs,dcolumn}
\usepackage{natmove}
\usepackage{adjustbox}
\usepackage{mciteplus}

\begin{document}

\title{An analysis of the performance of 
coupled cluster methods for core excitations and
core ionizations using standard basis sets}
\author{Johanna P. Carbone}
\affiliation{Dipartimento di Scienze Chimiche e Farmaceutiche, 
Universit\`a degli Studi di Trieste, I-34127, Trieste, Italy}
\author{Lan Cheng}
\affiliation{Department of Chemistry,  Krieger School of Arts and Sciences, Johns Hopkins University, Baltimore, MD 21218, USA}
\author{Rolf H. Myhre}
\affiliation{Department of Chemistry, Norwegian University of Science and Technology, N-7491 Trondheim, Norway}
\author{Devin Matthews}
\affiliation{Institute for Computational Engineering and Sciences, University of Texas at Austin, Austin, TX 78712, USA}
\author{Henrik Koch}
\affiliation{Department of Chemistry, Norwegian University of Science and Technology, N-7491 Trondheim, Norway}
\affiliation{Scuola Normale Superiore, Pisa, Italy}
\author{Sonia Coriani}\thanks{Corresponding author. Email: soco@kemi.dtu.dk}
\affiliation{Department of Chemistry, Technical University of Denmark, DK-2800 Kgs. Lyngby, Denmark}

\begin{abstract}
An extensive analysis has been carried out of the performance of standard families of basis sets with the hierarchy of coupled cluster methods CC2, CCSD, CC3 and CCSDT in computing selected Oxygen, Carbon and Nitrogen K-edge (vertical) core excitation and ionization energies within a core-valence separated scheme in the molecules water, ammonia, and carbon monoxide. Complete basis set limits for the excitation energies  have been estimated via different basis set extrapolation schemes.
The importance of scalar relativistic effects has been established within the spin-free exact two-component theory in its one-electron variant (SFX2C-1e).
\end{abstract}

\maketitle

\section{Introduction}
\label{Introduction}
Core-level spectroscopy, including techniques such as Near-Edge Absorption Fine Structure and X-ray Photoelectron spectroscopies, is widely used in various areas of contemporary research, such as in surface science,  organic electronics,  and medical biological research \cite{Bergmann:FEL:17}.
It is considered a powerful tool to gain insight into the electronic structure of molecular species. The recent improvements of the synchrotron radiation sources
and the emergence  of the free-electron laser  have further broadened the range of phenomena and systems that can be studied by core-level spectroscopy, see e.g.
Refs.
\cite{Bergmann:FEL:17,VanKuiken2012,piancastelli_perspective,Milne2014,Lamberti:XRay:2016,naturecomm,xray_rev2,Lin2017,kraus2018}.
An essential requirement for a successful application of core-level techniques is the
availability of reliable computational methods that allow for
a proper interpretation of the resulting spectra.
Several quantum chemical approaches exist
for the calculation of core-excited/ionized states. While referring to Ref.~\citenum{Norman:ChemRev:18} for a recent review,
we mention here as examples  the symmetry-adapted cluster configuration
interaction (SAC-CI)~\cite{SACCI}, the GW approximation
(self-energy approximated by Green-function G and screened Coulomb W)
to the Bethe--Salpeter equation~\cite{GWBS,GILMORE2015109},
the static-exchange (STEX) approach~\cite{STEX}, and the restricted and unrestricted algebraic diagrammatic construction scheme (ADC)
~\cite{ADC82,ADC_review,wenzel2014,wenzel2015}
up to third order exploiting the core--valence separation (CVS)~\cite{Cederbaum1980} approximation.
Large systems are often treated with
time-dependent density functional
theory (TD-DFT)~
\cite{Stener2003,Ekstrom:2006p2332,Ekstrom:2006p2752,Tu:2007p2336,REWTDDFT,Besley:2010p8661,REWTDDFT2},
but the results are plagued by the self-interaction error and the arbitrary dependence on the choice of the exchange-correlation functional.
Indeed, unless short-range corrected hybrid functionals are used~\cite{besley2009self},
core-excited states calculated by TD-DFT with conventional functionals often reproduce experimental spectra qualitatively well, but the self-interaction error and the small gap between occupied and unoccupied electronic levels inherent in the TD-DFT formalism lead to underestimation of core-excited states.
Therefore absolute
core excitation energies obtained by conventional TD-DFT
are typically corrected by shifting them tens of eVs in order
to agree with experiment.
Among the time-independent DFT-based approaches for core excitations, we mention the recently proposed, and remarkably accurate, variational orthogonality constrained density functional theory method of Evangelista and co-workers~\cite{derricotte2015simulation,evangelista2013},
see also the work of Glushkov, Assfeld and coworkers
on orthogonality constrained/local Hartree-Hock Self-Consistent-Field \cite{OC_HF_Glushkov_2016,OC_HF_Glushkov_2017,LocalSCF,LocalSCF2}.
Over the last eight years, we have made a significant effort to extend the applicability of the coupled cluster linear response
(CC-LR)~\cite{koch1990,christiansen1998ijqc}
and equation-of-motion coupled-cluster (EOM-CC)~\cite{stanton_equation_1993,bartlett_coupled-cluster_2012,krylov_eom_2008} formalisms to the computation
of core-level  spectroscopies~\cite{coriani2012pra,coriani2012jctc,fransson2013jcp,List2014,coriani2015jcp,
cvs:erratum,mlcc_cvs,naturecomm,myhre2018N2,
marta2019,lan2019,rasmus2019,federica2019}.
The CC ansatz is known to provide a systematic hierarchy of models with increasing accuracy, allowing for the prediction of molecular properties and spectra with controlled accuracy within the hierarchy \cite{WavFuncRev,christiansen1998ijqc,helgaker2004}.

With the introduction in 2015 of CVS and
restricted-excitation-window schemes within CC-LR and EOM-CC~\cite{coriani2015jcp,cvs:erratum,Peng2015},
the use of CC methods for the determination of core-absorption spectra and core ionization energies has become as straightforward as it is for UV-vis excitations.
Since the computational determination of spectroscopic observables related to the interaction of the sample  with X-ray radiation displays strong dependence on the level of theory and size of the basis sets, a systematic approach becomes particularly attractive.
An important component to this end is a rigorous assessment of the basis set requirements and the relative accuracy of the various CC approximations, when computing core spectra using the different members of the CC(-LR) hierarchy.
This study is meant as a contribution in this direction, in the spirit of a similar study conducted within the ADC formalism~\cite{wenzel2015}.

\section{Methodology and Computational Details}

Calculations of core excitation energies,  oscillator strengths (in length gauge) and ionization energies (IE) have been performed for the  hierarchy of CC methods: coupled cluster singles and approximate doubles (CC2), coupled cluster singles and doubles (CCSD), coupled cluster singles, doubles and approximate triples (CC3), and  coupled cluster singles, doubles and triples (CCSDT).
All methods are extensively described in the literature,
see e.g. Refs.\citenum{CC2,CCSD,directCCSD2,singletexci,CC3,CC3resp,Kucharski2001},
and we refrain therefore from repeating their derivation here.
We limit ourselves to draw the reader's attention to Ref.~\citenum{coriani2015jcp} for
the description of how the core-valence separation scheme used here has been implemented within CC-LR/CC-EOM, to Ref.~\citenum{mlcc3} for a new, more efficient, implementation of CC3, and to Refs.~\citenum{CCSDTQ1,CCSDTQ2} for efficient implementation of the CCSDT method.
Scalar-relativistic effects have been taken into account in the CCSDT 
using  the spin-free exact two-component theory in its one-electron variant (SFX2C-1e)\cite{SFX2C1,SFX2C2,SFX2C3}.

All calculations up to CC3 have been run with a development version of the
Dalton code~\cite{DaltonPaper}, whereas the
CFOUR~\cite{CFOUR} code  was used for the CCSDT
core excitation and ionization energies, respectively.
Accurate experimental equilibrium geometries were adopted for all three systems: $R_e$ (CO) = 1.12832  {\AA} for CO; $R_e$ (NH) = 1.011 {\AA} and $\alpha_{{\rm HNH}}$ = 106.7$^{\circ}$ for ammonia; $R_e$ (OH) = 0.9570 {\AA} and $\alpha_{\rm{ HOH}}$ = 104.5$^{\circ}$ for water.

\section{Results and discussion}

\subsection{Excitation energies}
\label{energies}

\begin{figure}
\caption{H$_2$O, Oxygen K-edge. Basis set convergence of the first
two vertical core excitation energies with different CC methods and basis sets.}
\label{H2O_2states}
\vspace*{-1.0cm}
\begin{center}
\hspace*{-1.8cm}\includegraphics[scale=0.45]{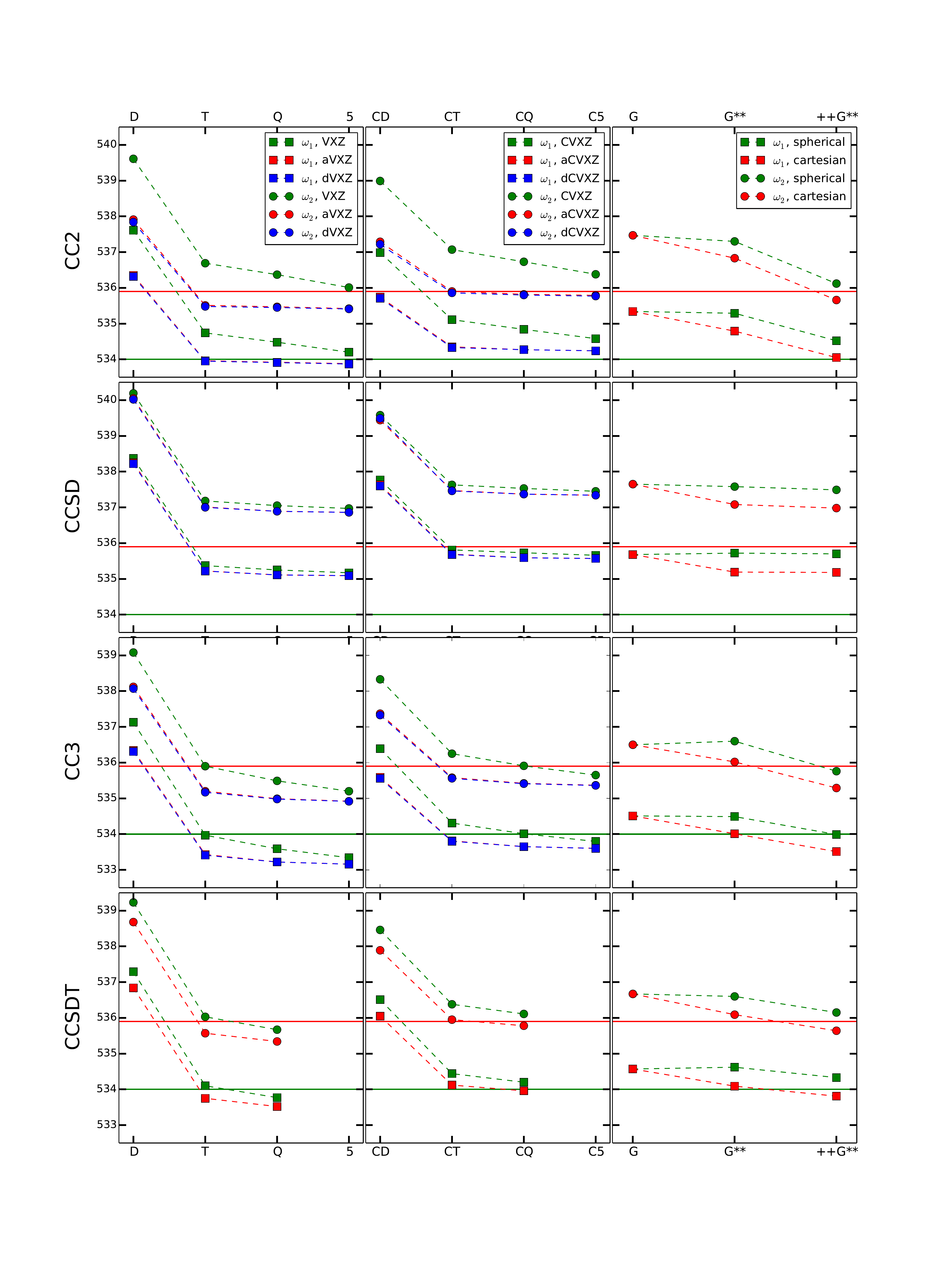}
\end{center}
\end{figure}

\begin{figure}
\caption{CO, Carbon K-edge. Basis set convergence of the first two vertical core excitation energies  with different CC methods and basis sets.}
\label{CO_C_2states}
\vspace*{-1.0cm}
\begin{center}
\hspace*{-1.8cm}\includegraphics[scale=0.45]{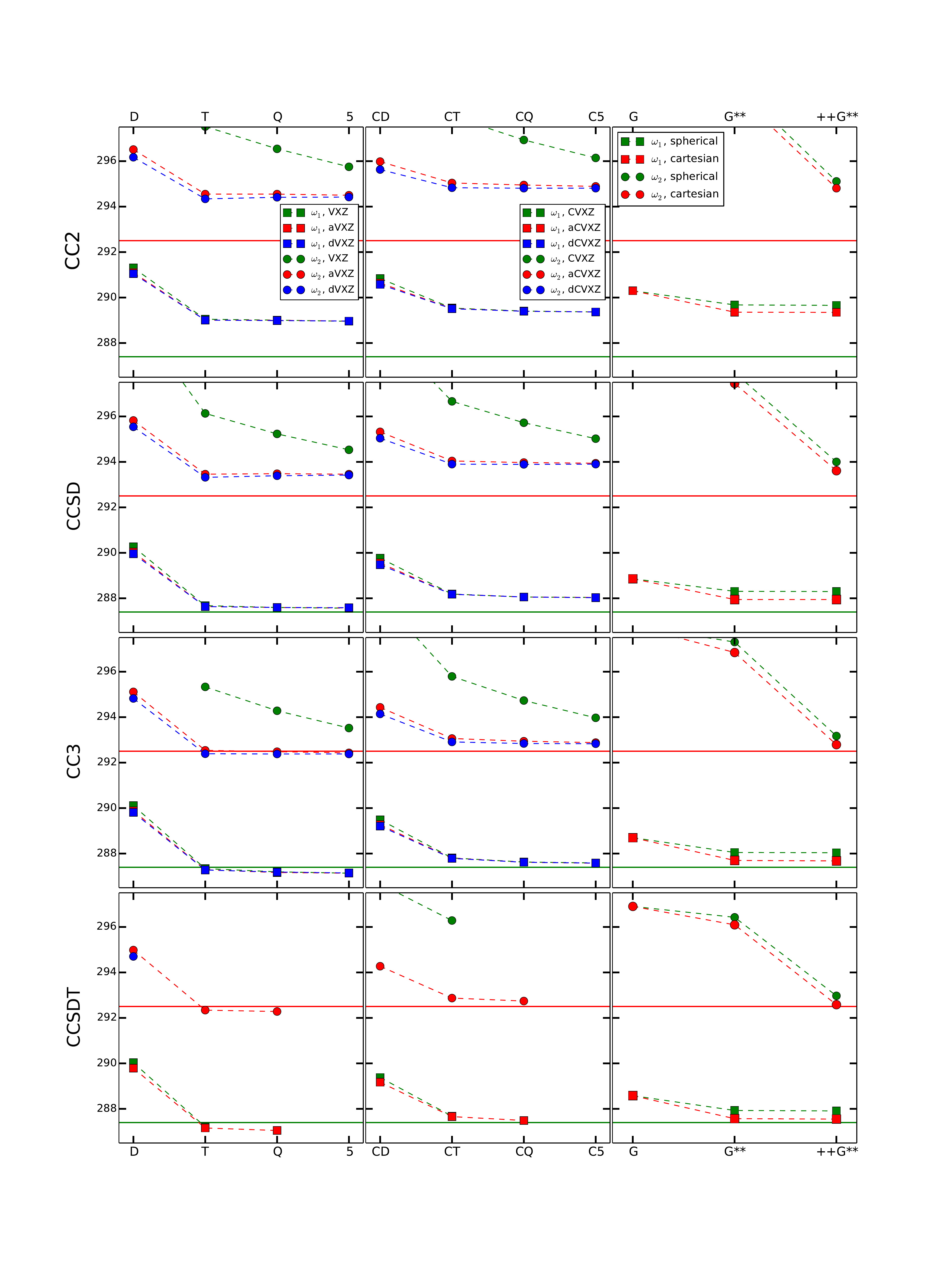}
\end{center}
\end{figure}

\begin{figure}
\caption{CO, Oxygen K-edge. Basis set convergence of the first two vertical core excitation energies  with different CC methods and basis sets.}
\label{CO_O_2states}
\vspace*{-1.0cm}
\begin{center}
\hspace*{-1.8cm}\includegraphics[scale=0.45]{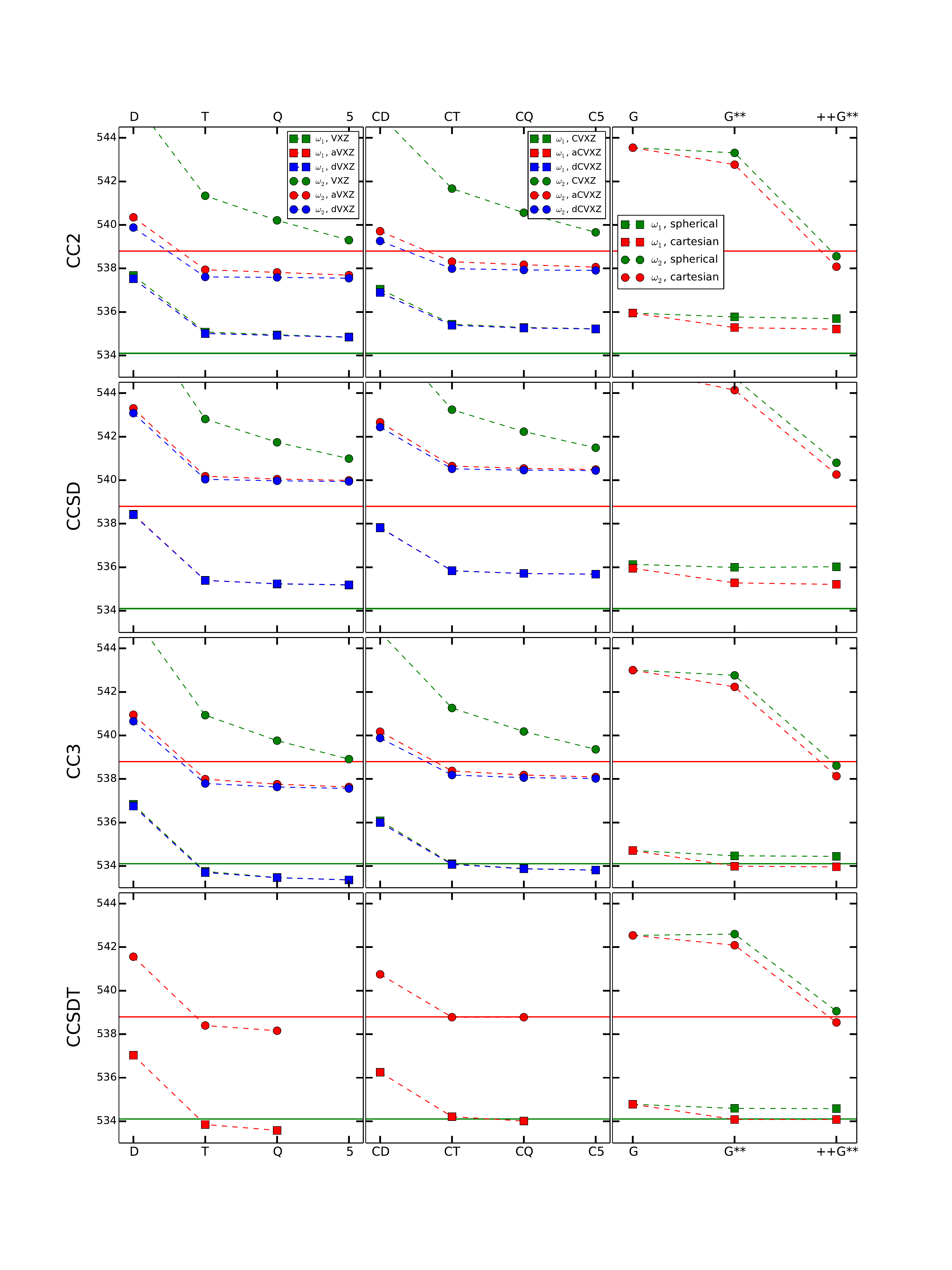}
\end{center}
\end{figure}

\begin{figure}
\caption{NH$_3$, Nitrogen K-edge. Basis set convergence of the first two vertical
core excitation energies  with different CC methods and basis sets.}
\label{NH3_2states}
\vspace*{-1.0cm}
\begin{center}
\hspace*{-1.8cm}
\includegraphics[scale=0.45]{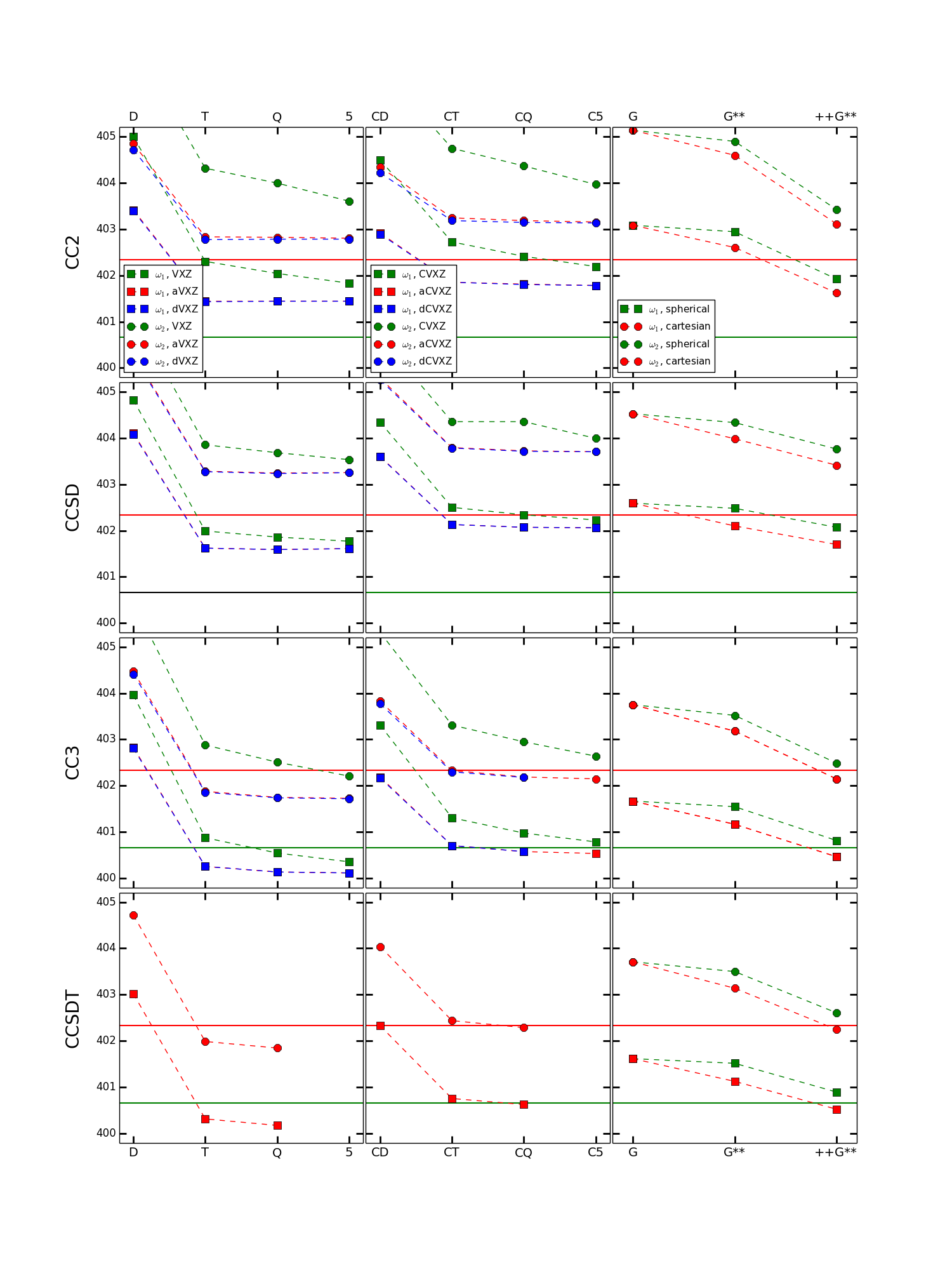}
\end{center}
\end{figure}

We start our discussion with a detailed comparison of the
computed excitation energy values using the different basis sets
across the CC hierarchy and
with respect to the experimentally derived values.
To this end, we plot
in Figs.~\ref{H2O_2states}
to~\ref{NH3_2states} the trends within
each basis set family and method for the considered excitation energies.
We will focus our discussion primarily on core excitations that are
individually resolved in the experimental spectra:
the 1s$\to$3s (a$_1$) and 1s$\to$3p (b$_2$) transitions in water,
the 1s$\to\!\pi^*$ transition for both C and O of CO,
and the first three core excitations in NH$_3$.
The third intense peak in the X-ray absorption spectrum of H$_2$O
is known to originate from the overlap of core transition into
a$_1$ and b$_1$ states and will be commented upon in Section~\ref{spectra}.
The full set of numerical values of the core-excitation energies
is available on arXiv~\cite{Tables:AQC}.

Within each series of correlation consistent (cc) basis sets (regular, single augmented and double augmented), we observe an almost monotonically-decreasing trend (towards the experimental value) while  increasing the
basis set cardinal number.

In the cc-pVDZ basis, the excitation energies are always overestimated, by 2 to 5 eV depending on the case, with respect to both the other members of the series and the experimental values. By further increasing the cardinal number, the differences within each series reduce to tenths or hundredths of an eV.
In other words, any Dunning set of X$\ge$3 is reasonably accurate, and the results are significantly improved by inclusion of the first level of augmentation. Double augmentation has moderate effects for the chosen core excitations.

For the 1s$\rightarrow$3s(a$_1$) in water, for instance,
the differences between the CCSD results
obtained for X=2 (DZ) and X=3 (TZ) in the series
cc-pVXZ are always of the order of 3 eV, and slightly lower than  2~eV for the cc-pCVXZ series, progressively reducing to tenths and hundredths of an eV when increasing X.
With reference to the experimental value for the same excitation,
the basis sets with X=2 overestimate the edge by ca. 3.5--4.5 eV (depending on the basis); for X$\geq$3, the deviation is reduced to ca. 1.0--1.2 eV for the
(x-aug-)cc-pVXZ sets, and to  ca. 1.5--1.7 eV for the (singly and doubly augmented) cc-pCVXZ sets.

The trend observed for the first excitation is roughly the same also for the second one.  However, for the third and fourth core excitations of water (third peak in the
experimental spectrum), as well as any higher lying excitations of more diffuse/Rydberg character than those considered here, it becomes of paramount importance to include additional diffuse functions~\cite{coriani2012pra,marta2019,federica2019}.

Among the Pople basis sets, the 6-311++G** set emerges as remarkably accurate in basically all cases (states and methods) despite its moderate size, as also previously observed for the ADC family of methods~\cite{wenzel2014}. Use of Cartesian $d$ functions is  to be slightly preferred.

The CCSD model systematically overestimates all core excitation energies (roughly of the same amount for all excitations), allowing for a 'rigid--shift' correction.
The CC2 core excitation energies tend to be smaller than the CCSD ones, and
they can be both red-shifted and blue-shifted compared to their experimental counterparts.
For the first excitation, they are, at first sight, also closer to the experimental value,
but the peak separation is underestimated. As we will see in Section~\ref{spectra}, this, together with the results for intensities, actually results in a poor comparison of the CC2 spectral profile with the experimental one, at least for the three systems considered here.

\subsection{Extrapolation towards the complete basis set (CBS) limits}
\label{limits}

As observed in Section~\ref{energies}, the results in the cc basis sets show a monotonically decreasing behaviour when increasing the cardinal number. The cc basis sets are known to yield a systematic convergence towards the complete basis set limit for the correlation energy of the ground state, as well as for other molecular properties, and various extrapolation formulae have been proposed in the literature.
Some of these formulae tend to overestimate the limit,
and others to underestimate it. Inspired by the analysis performed by \citeauthor{wenzel2014} for the ADC hierarchies \cite{wenzel2014},
we have considered whether two popular extrapolations formulas,
namely the X$^{-3}$  formula~\cite{HelgakerExtra} and the exponential formula~\cite{exponential_formula},
\begin{align}
\label{extrap_formula}
E_{\rm{X}}&=E_{\rm{CBS}}+A {\rm{X}}^{-3}; \\
\label{extrap_formula2}
E_{\rm{X}}&= E_{\rm{CBS}}+A e^{-({\rm{X}}-1)}+B e^{-({\rm{X}}-1)^2}
\end{align}
can be applied to obtain an estimate of the CBS values of the core excitation
energies considered in this study.
In Eqs.~\ref{extrap_formula} and~\ref{extrap_formula2}, $E_{\rm{CBS}}$ is the resulting estimated energy of the CBS limit,
and $E_{\rm{X}}$ is the calculated energy using the basis with
the cardinal number X.

The two formulas have been applied in different ways in the literature for different properties.
One can fit directly the results of each basis set series, imposing the functional
forms in Eq.~\ref{extrap_formula}.
Alternatively, one can derive the CBS limits via either a two point strategy (on the X$^{-3}$ formula)  or three point  strategy (on the exponential formula), using the
energy values relative to the two (three) highest values of the cardinal numbers: X = Q, 5 for the two-point extrapolation, and  \mbox{X=T, Q, 5} for the three point extrapolation.
In the following, we have considered both strategies.

Notice that in standard basis set extrapolation schemes~\cite{HelgakerExtra,exponential_formula}, the exponential and X$^{-3}$ formulae apply to Hartree-Fock and correlation energies, respectively. Since a separation of the excitation energies into HF and correlation contributions are not straightforward, we apply the formulae directly to the computed excitation energies. It is also important to bear in mind that these extrapolation formulae are not rigorous expressions for the basis set dependence of energies, but serve as an estimate of the trend.

%

\begin{figure}[hbt!]
\caption{H$_2$O, Oxygen K-edge. Extrapolated complete-basis set (CBS) limits for the first two core excitations with different CC methods and the aug-cc-pCVXZ series. Upper panels: CC2; middle panels: CCSD; Bottom panels: CC3. The horizontal green lines correspond to the experimentally estimated core energies.}
\label{H2O_extrapolations}
\begin{center}
\vspace*{-0.5cm}
\hspace*{-1.0cm}\includegraphics[scale=0.4]{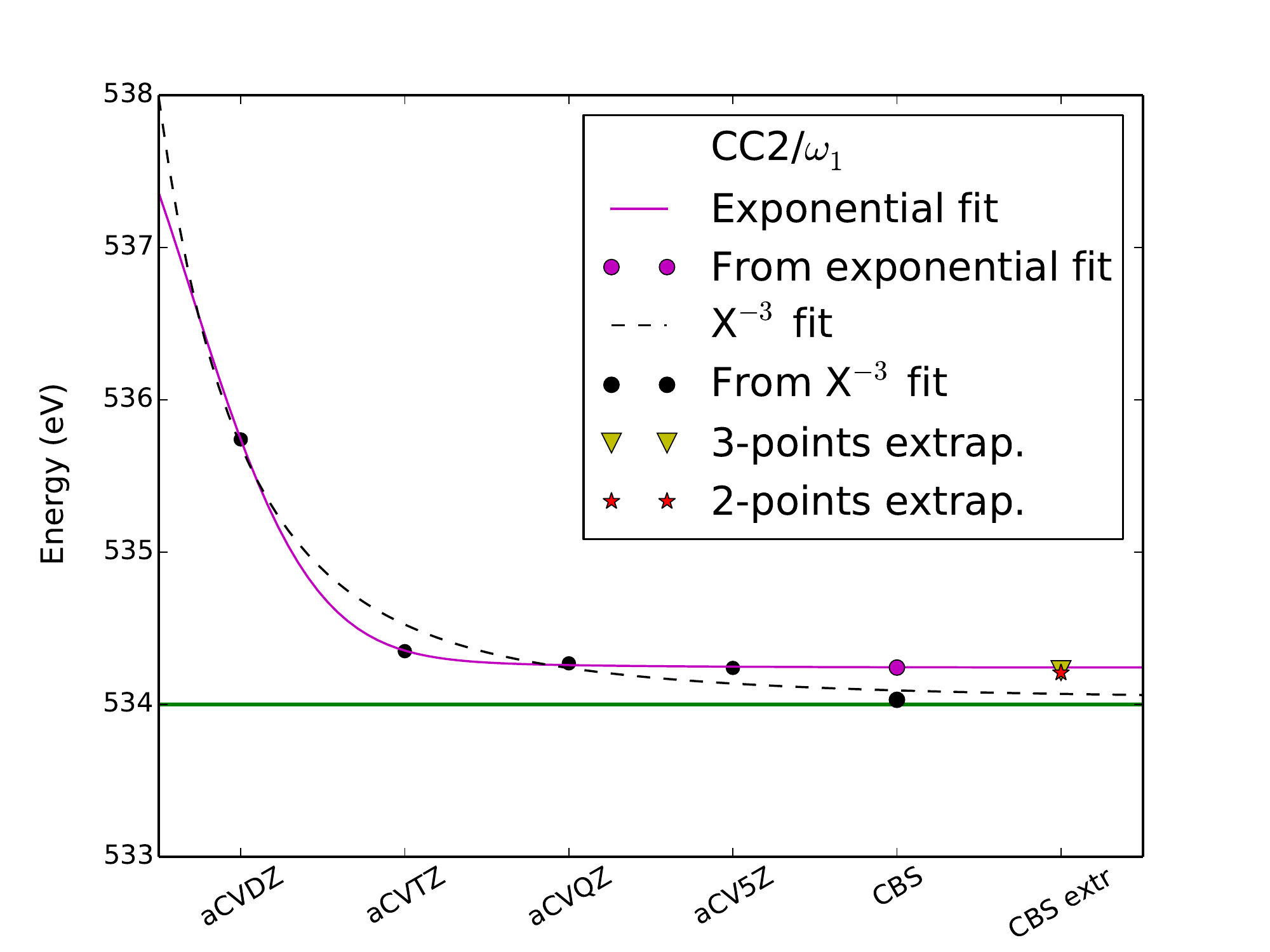}\hspace*{-0.3cm}\includegraphics[scale=0.4]{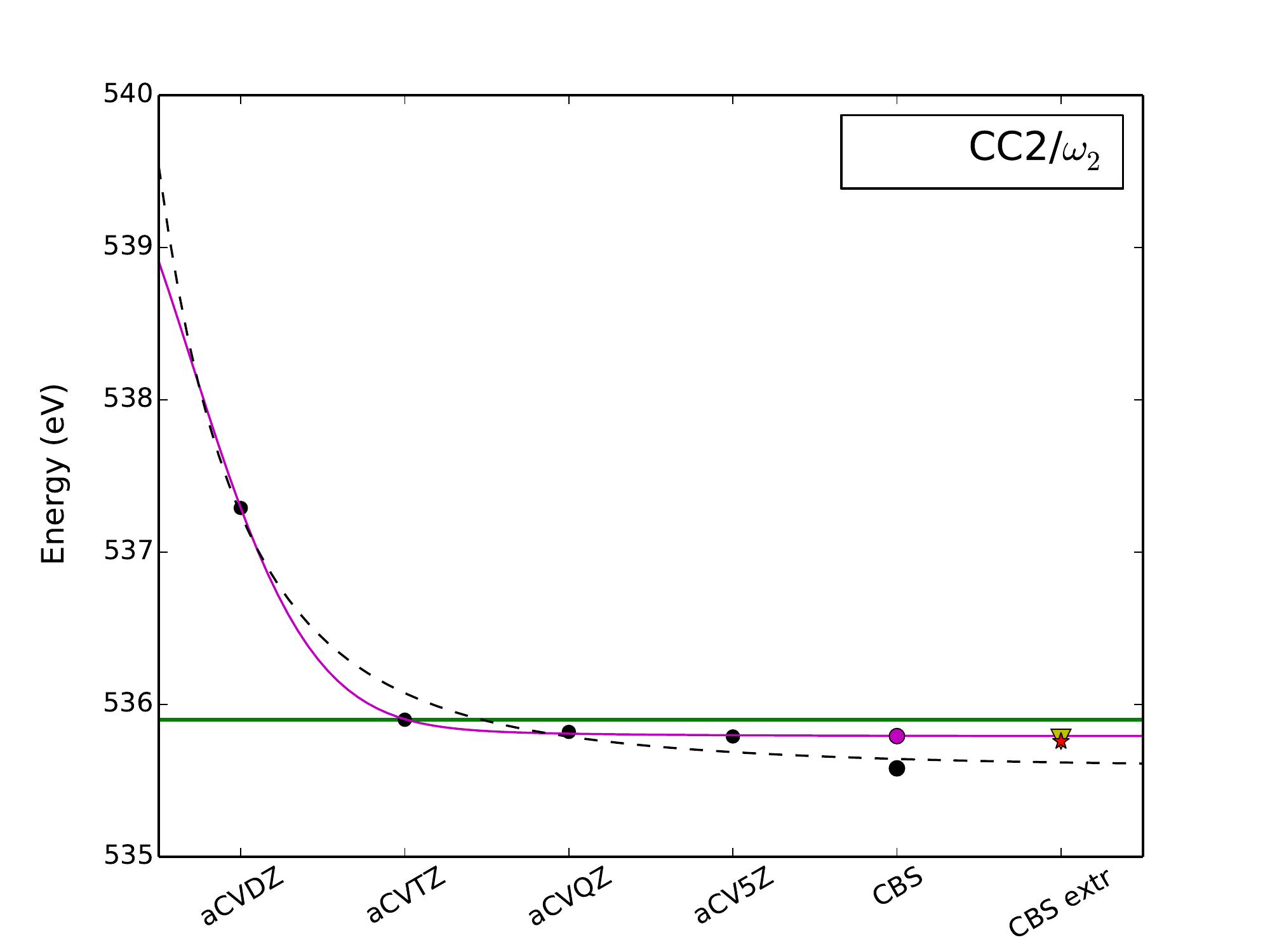}\\
\hspace*{-1.0cm}\includegraphics[scale=0.4]{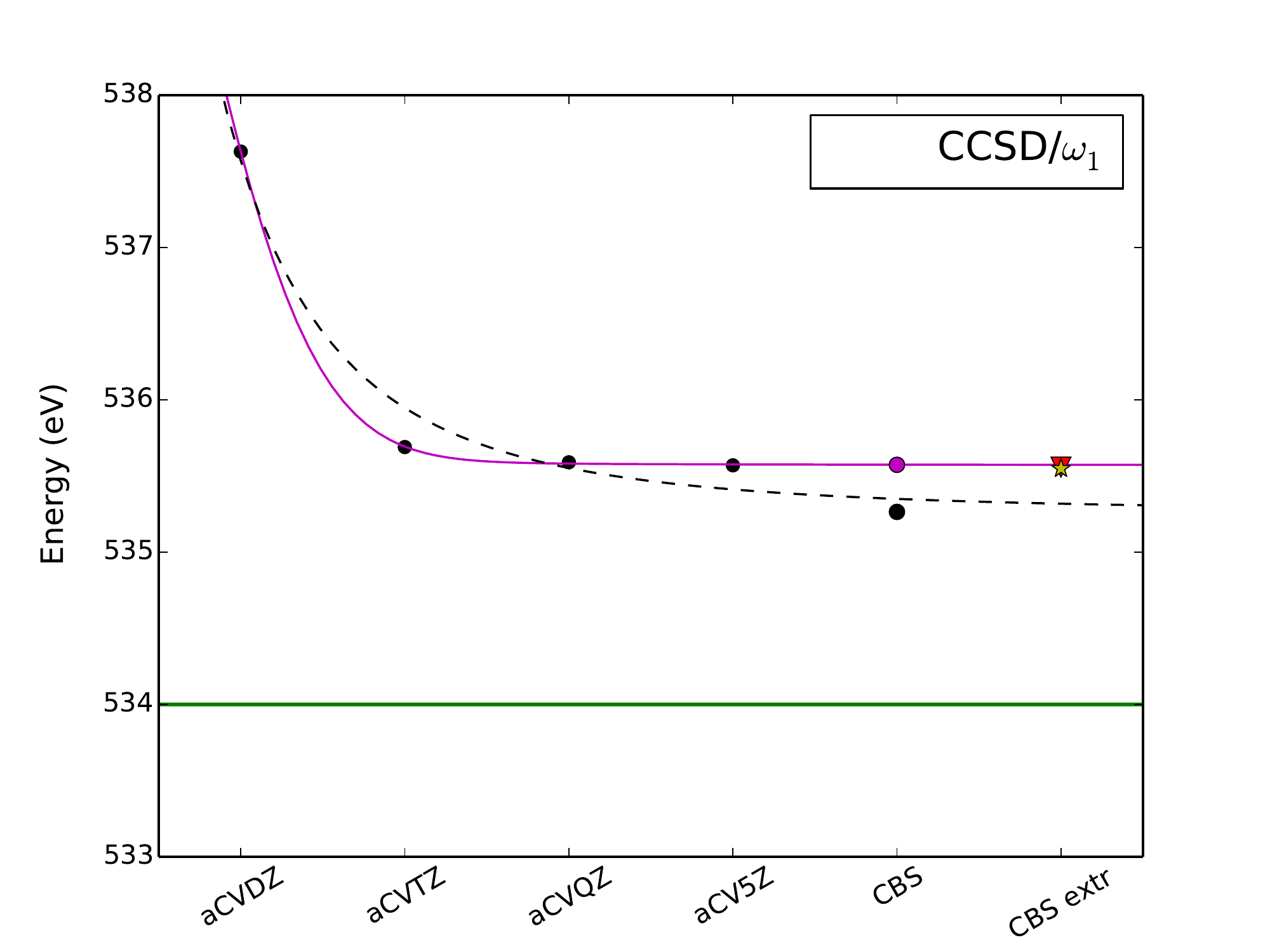}\hspace*{-0.3cm}\includegraphics[scale=0.4]{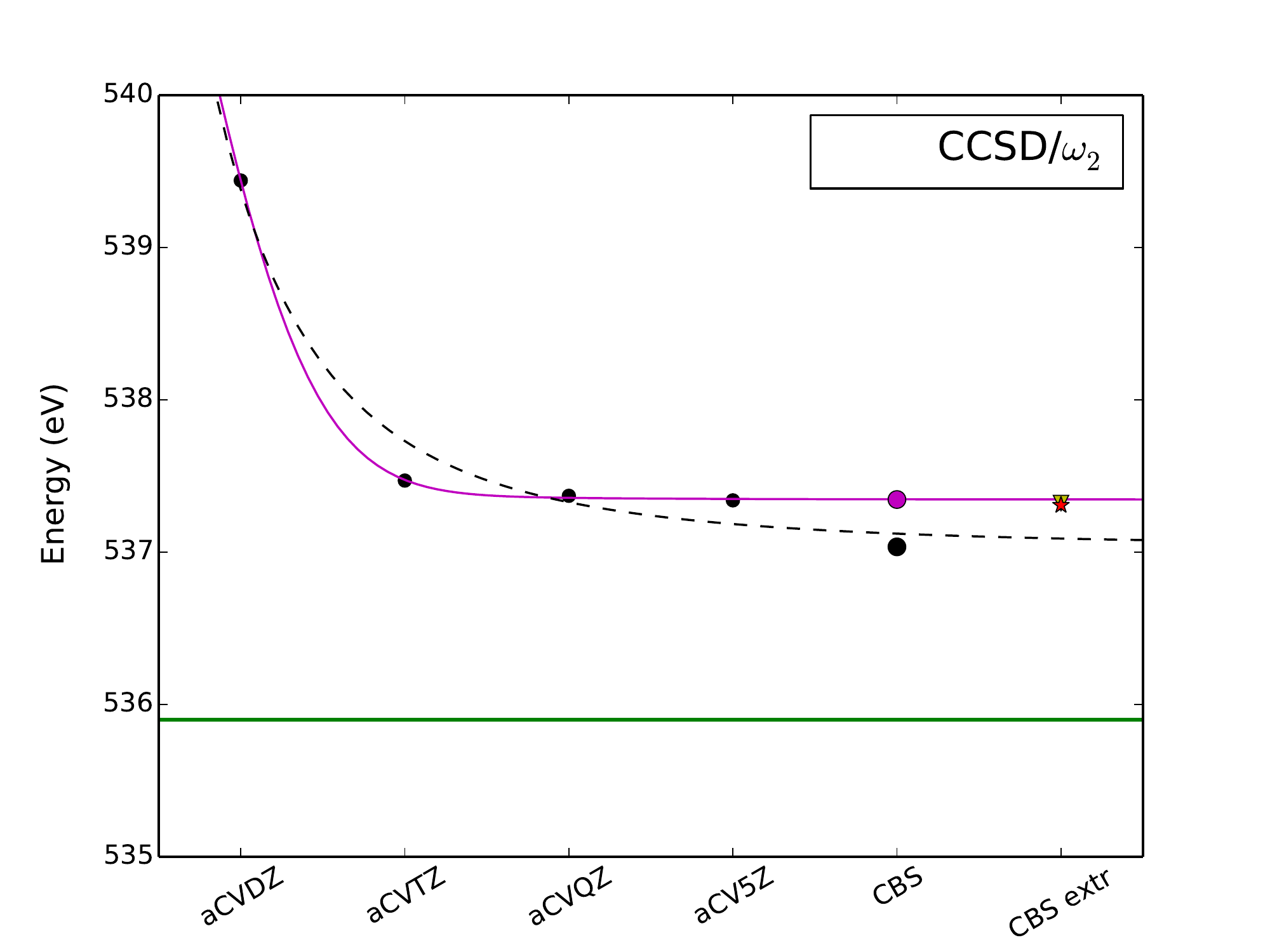}\\
\hspace*{-1.0cm}\includegraphics[scale=0.4]{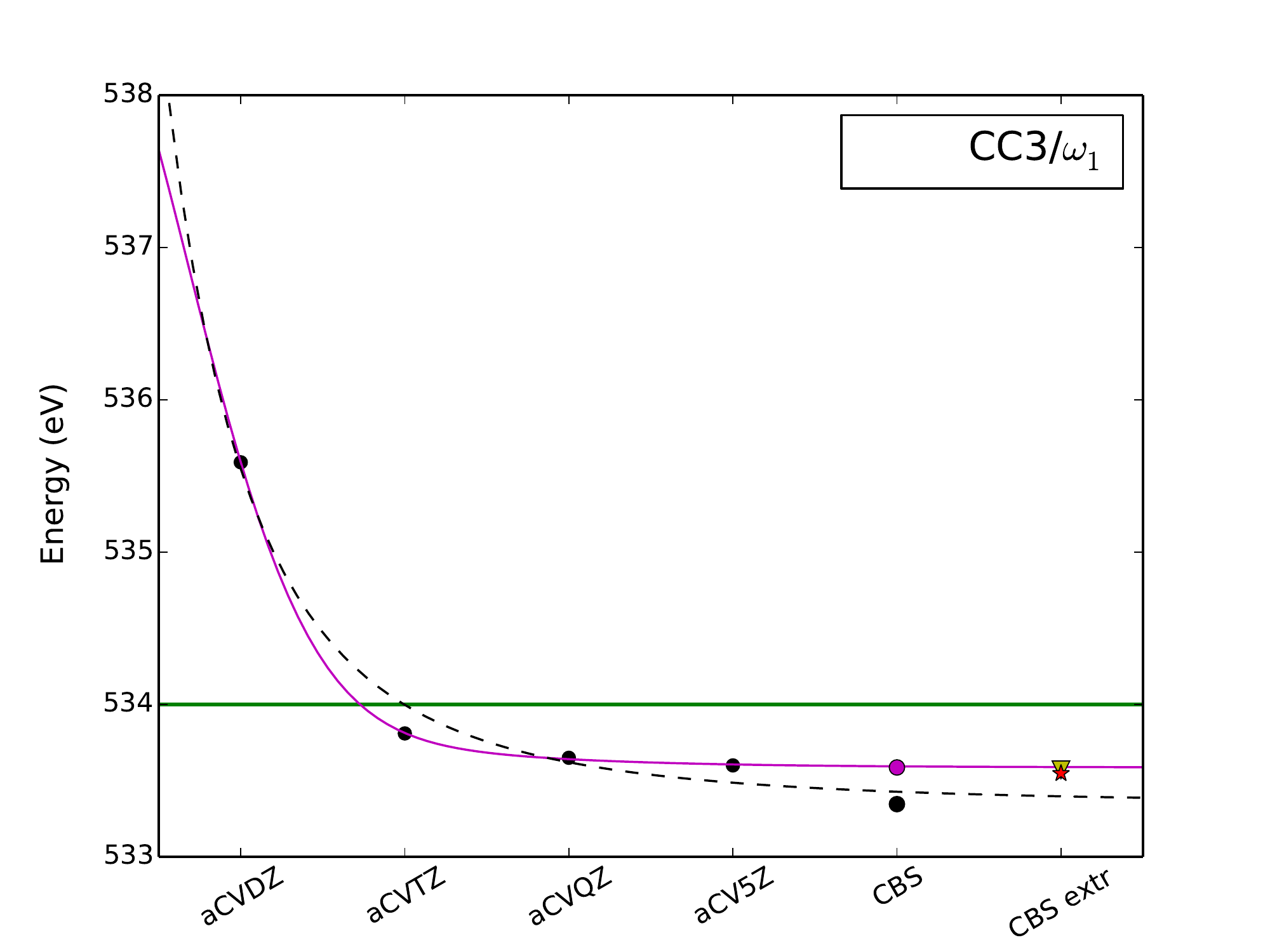}\hspace*{-0.3cm}\includegraphics[scale=0.4]{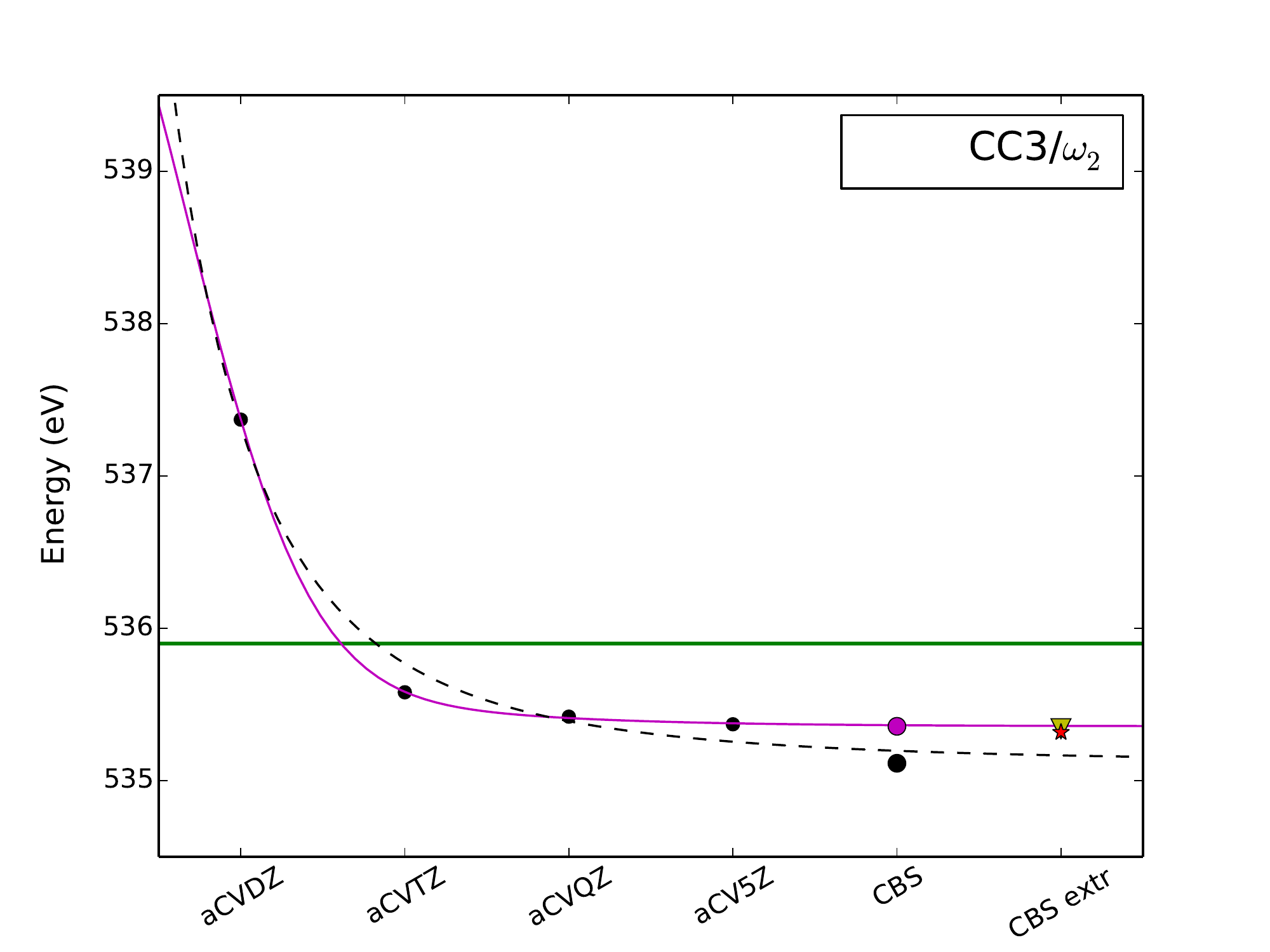}
\end{center}
\end{figure}

In Fig. \ref{H2O_extrapolations} 
we show the results obtained for  one selected basis set family, the aug-cc-pCVXZ one, for all four CC methods in the case of the
water molecule. The trends observed for CO (both edges) and NH$_3$ are completely analogous
and can be found in the document on the arXiv~\cite{Tables:AQC}.
The CBS values obtained directly from the two points X$^{-3}$ or three points exponential
procedures are basically identical, and only marginally different from those obtained
by fitting with the exponential regression over the entire series.
This difference is slightly larger than what was observed by \citeauthor{wenzel2014}~\cite{wenzel2014} for the ADC methods.

By fitting the results with a  X$^{-3}$ formula, on the other hand, we could not reproduce the behaviour of the excitation energies.

\subsection{Spectral bands}
\label{spectra}

For comparison and assignment of the experimental spectra, the intensities of the absorption bands are required, and they are here obtained from the computed oscillator strengths.  The full set of oscillator strengths obtained for the different basis sets at the CC2 and CCSD levels are available on the arXiv.

%

\begin{figure}[hbt!]
\caption{H$_2$O, Oxygen K-edge. CC2 (left panels) and CCSD (right panels) spectral profiles with the x-aug-CVXZ basis sets.
The augmented bases results are compared with the experimental spectrum~\cite{SchirmerExp}, which has been shifted and rescaled to roughly overlap with the first computed band (X=T,Q,5).}
\label{H2O_spectra}
\begin{center}
\hspace*{-1.0cm}\includegraphics[scale=0.4]{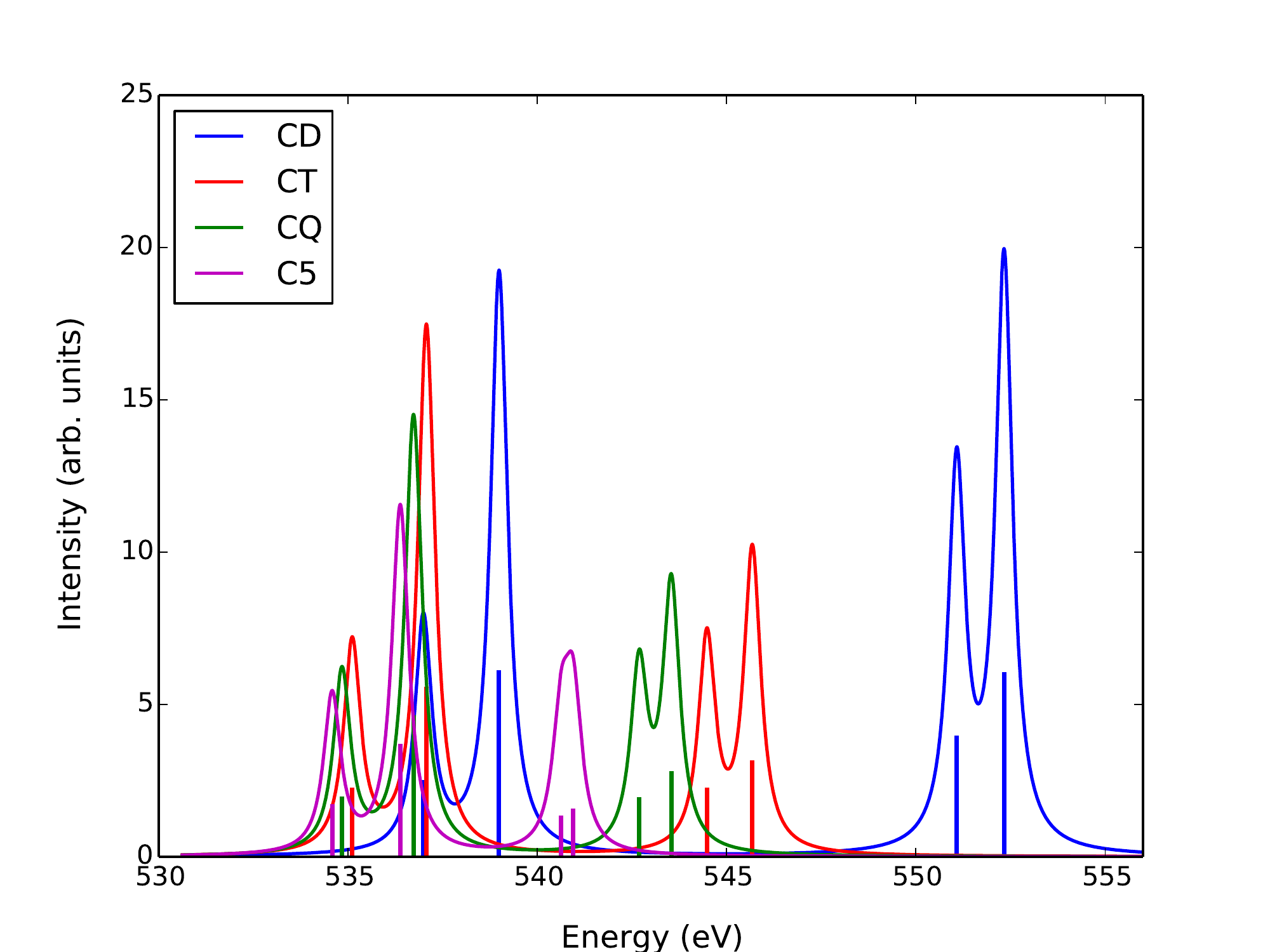}\hspace*{-0.5cm}\includegraphics[scale=0.4]{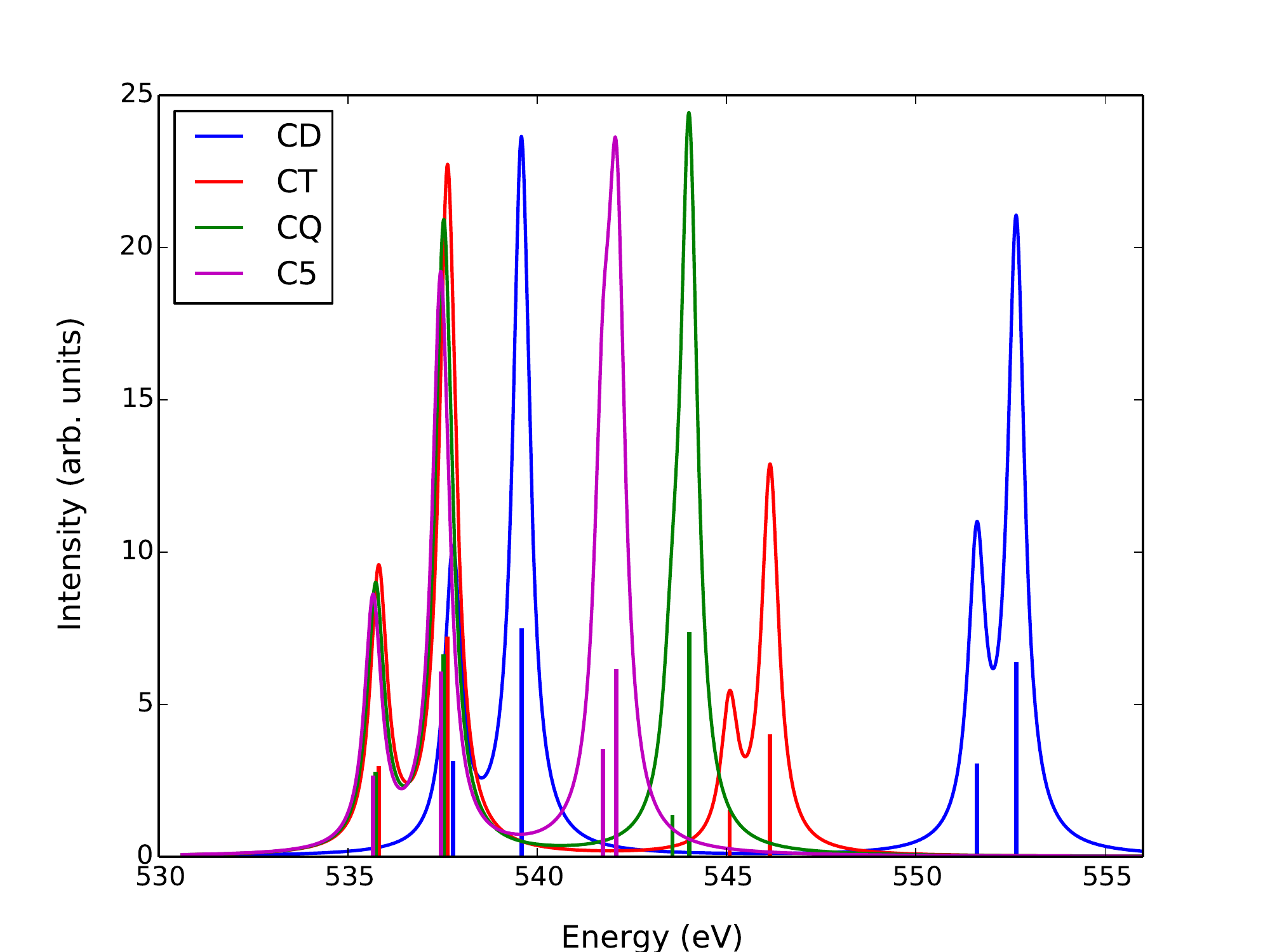}\\
\hspace*{-1.0cm}\includegraphics[scale=0.4]{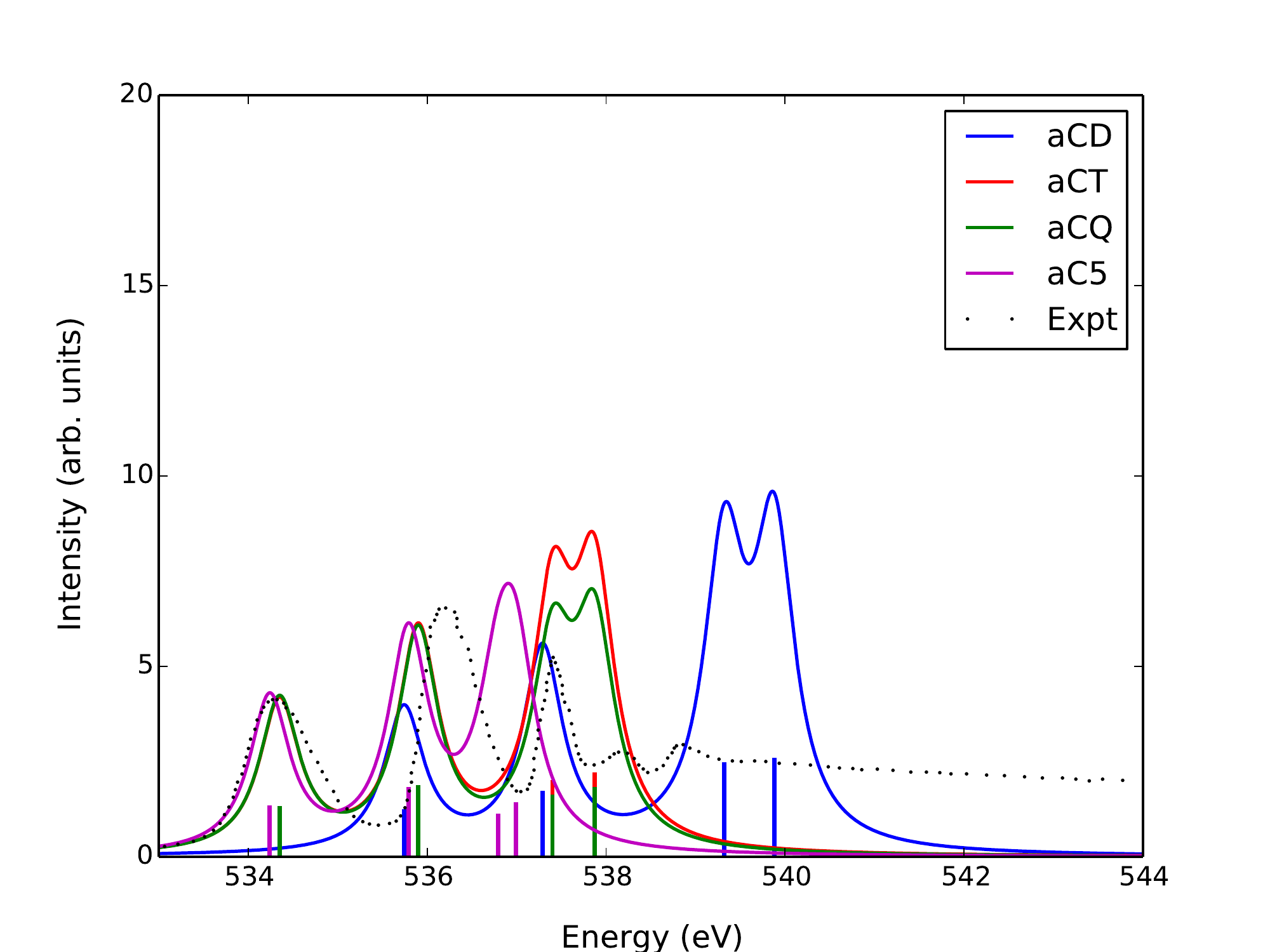}\hspace*{-0.5cm}\includegraphics[scale=0.4]{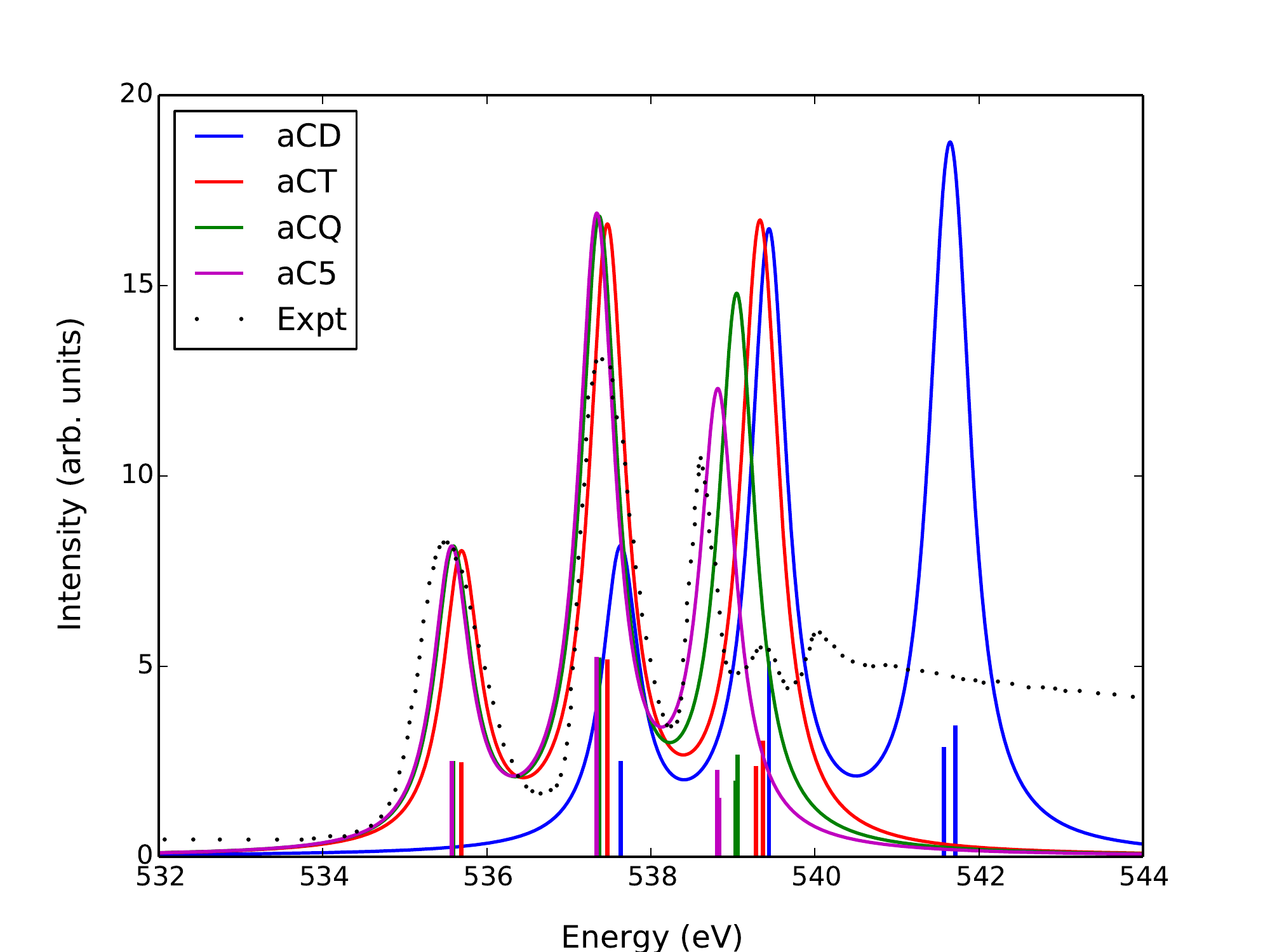}\\
\hspace*{-1.0cm}\includegraphics[scale=0.4]{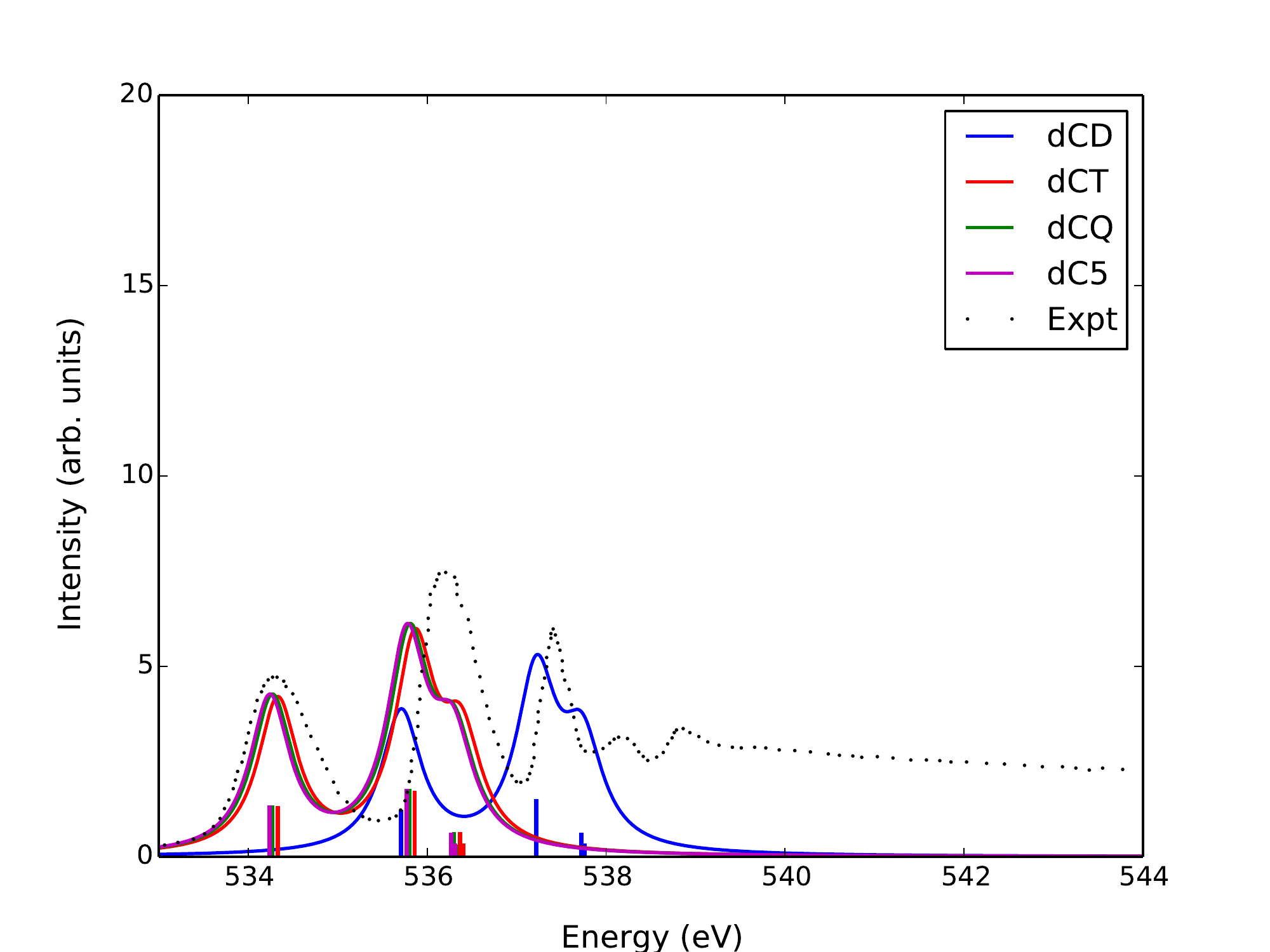}\hspace*{-0.5cm}\includegraphics[scale=0.4]{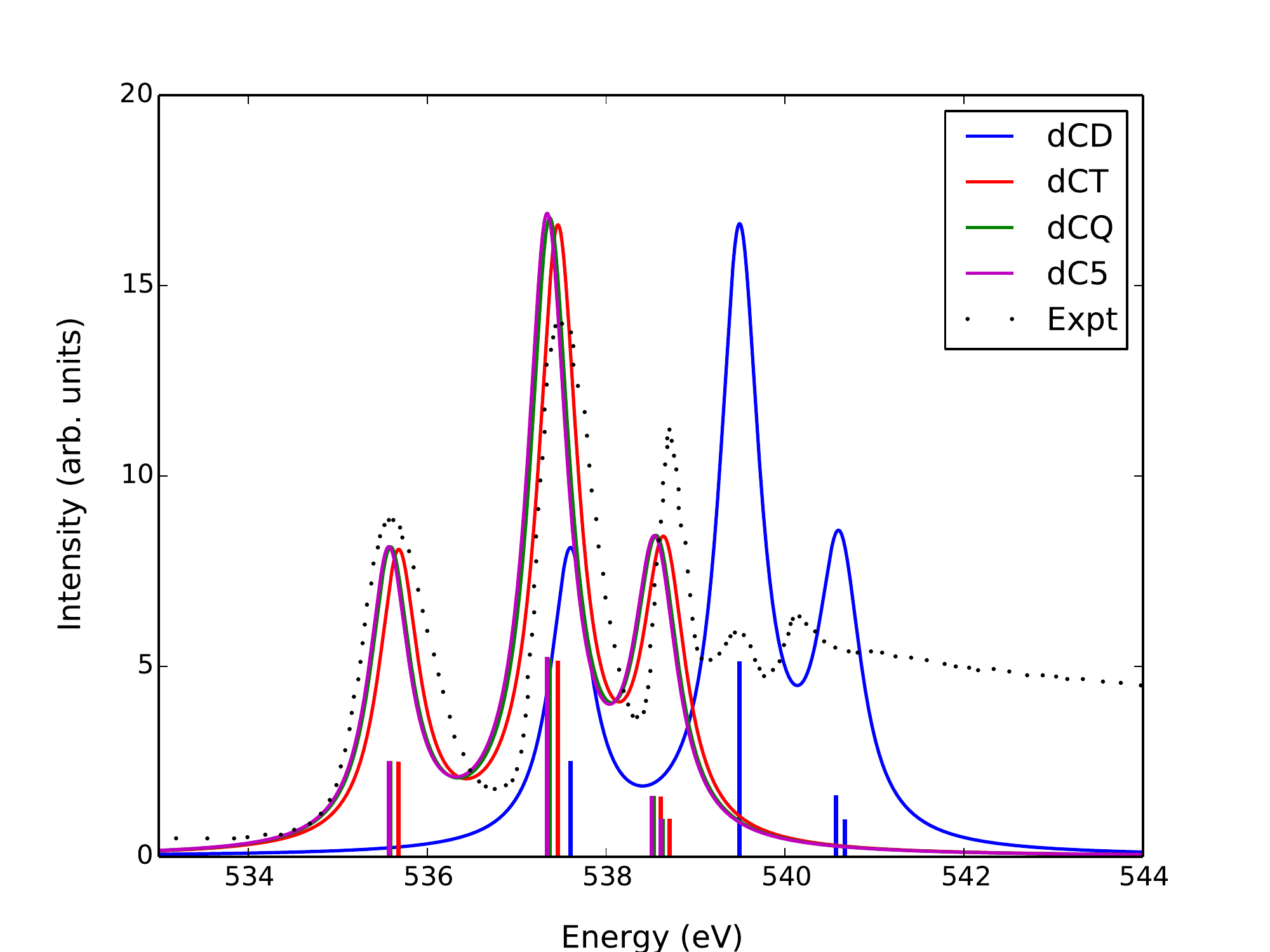}
\end{center}
\end{figure}

\begin{figure}[hbt!]
\caption{NH$_3$, Nitrogen K-edge. Spectral profiles (first 3 bands) at the CC2 (left) and CCSD (right) level in the cc-pCVXZ, aug-cc-pCVXZ and d-aug-cc-pCVXZ basis sets.
The augmented bases results are compared with the experimental spectrum~\cite{SchirmerExp}, which has been shifted and rescaled to roughly overlap with the first computed band (X$>$D).}
\label{NH3_spectra}
\begin{center}
\hspace*{-1.0cm}\includegraphics[scale=0.4]{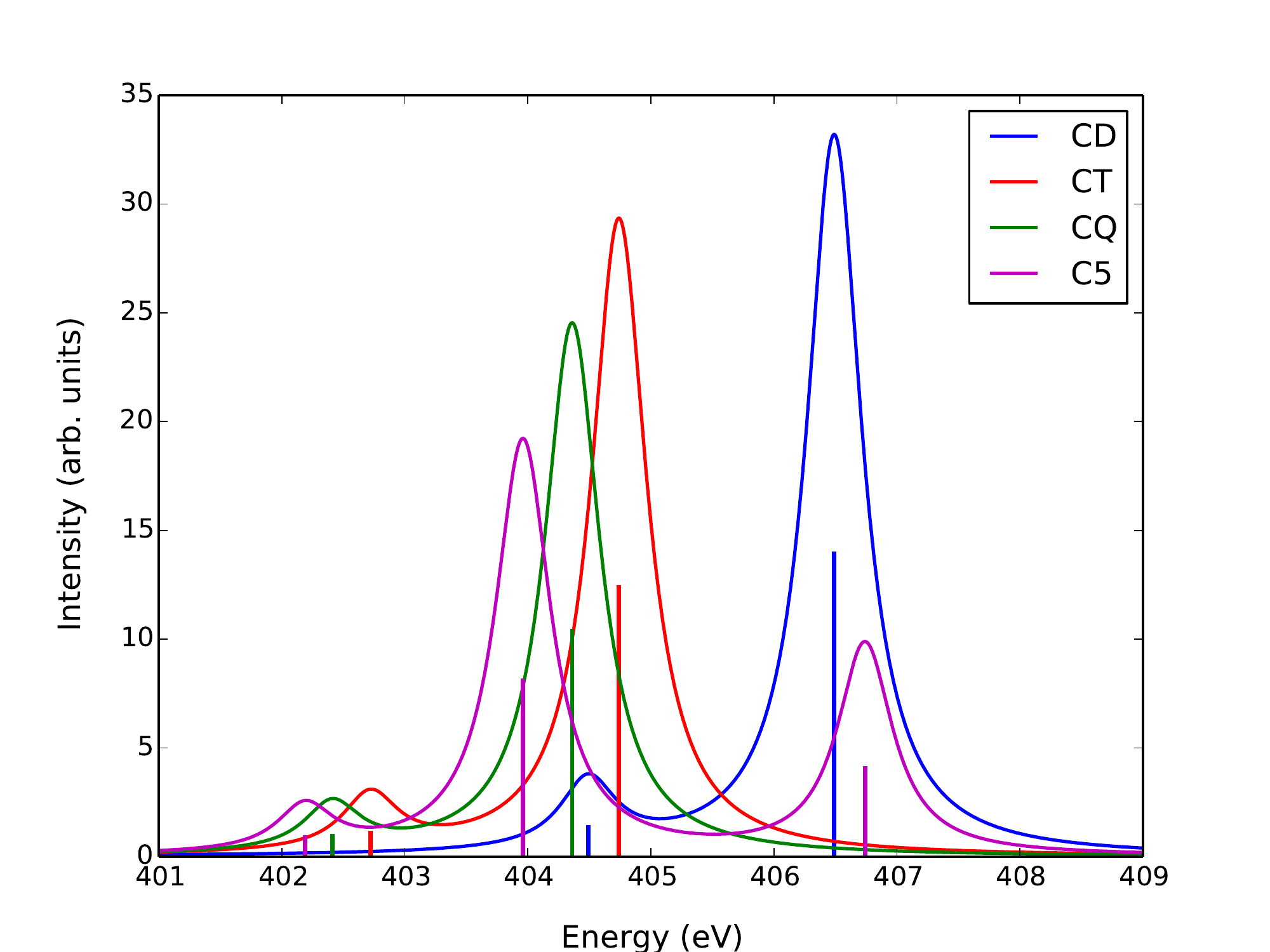}\hspace*{-0.5cm}\includegraphics[scale=0.4]{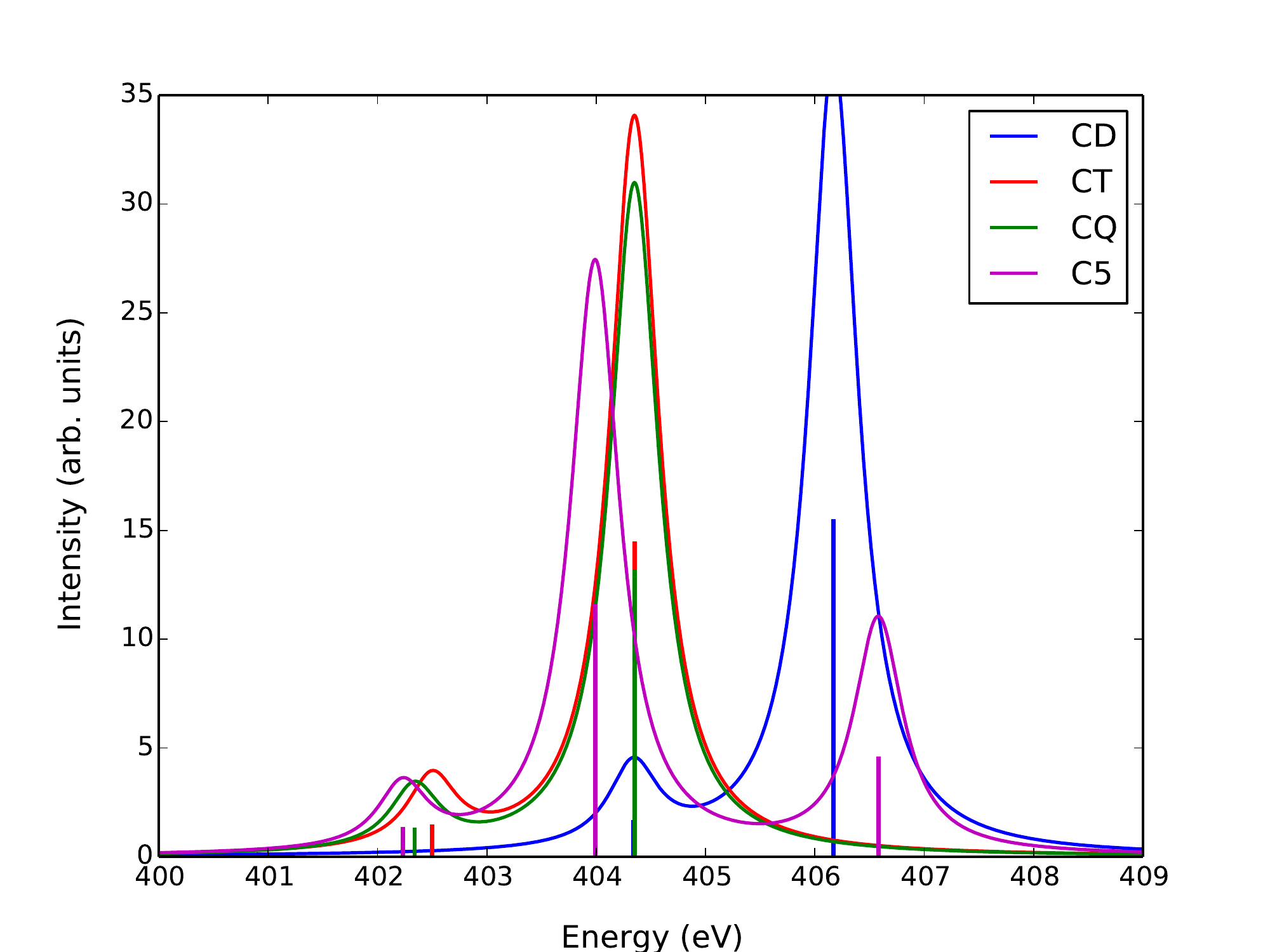}\\
\hspace*{-1.0cm}\includegraphics[scale=0.4]{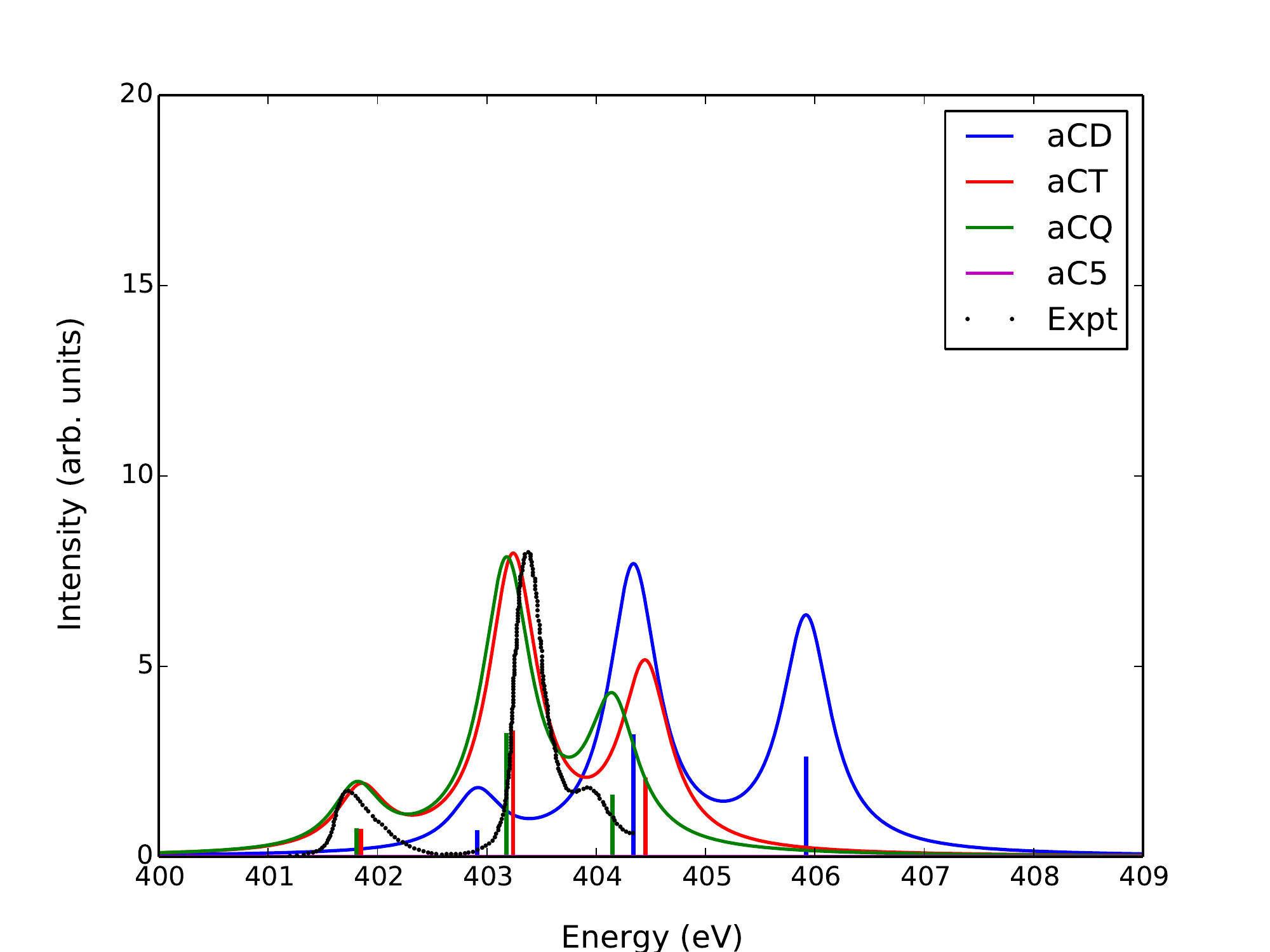}\hspace*{-0.5cm}\includegraphics[scale=0.4]{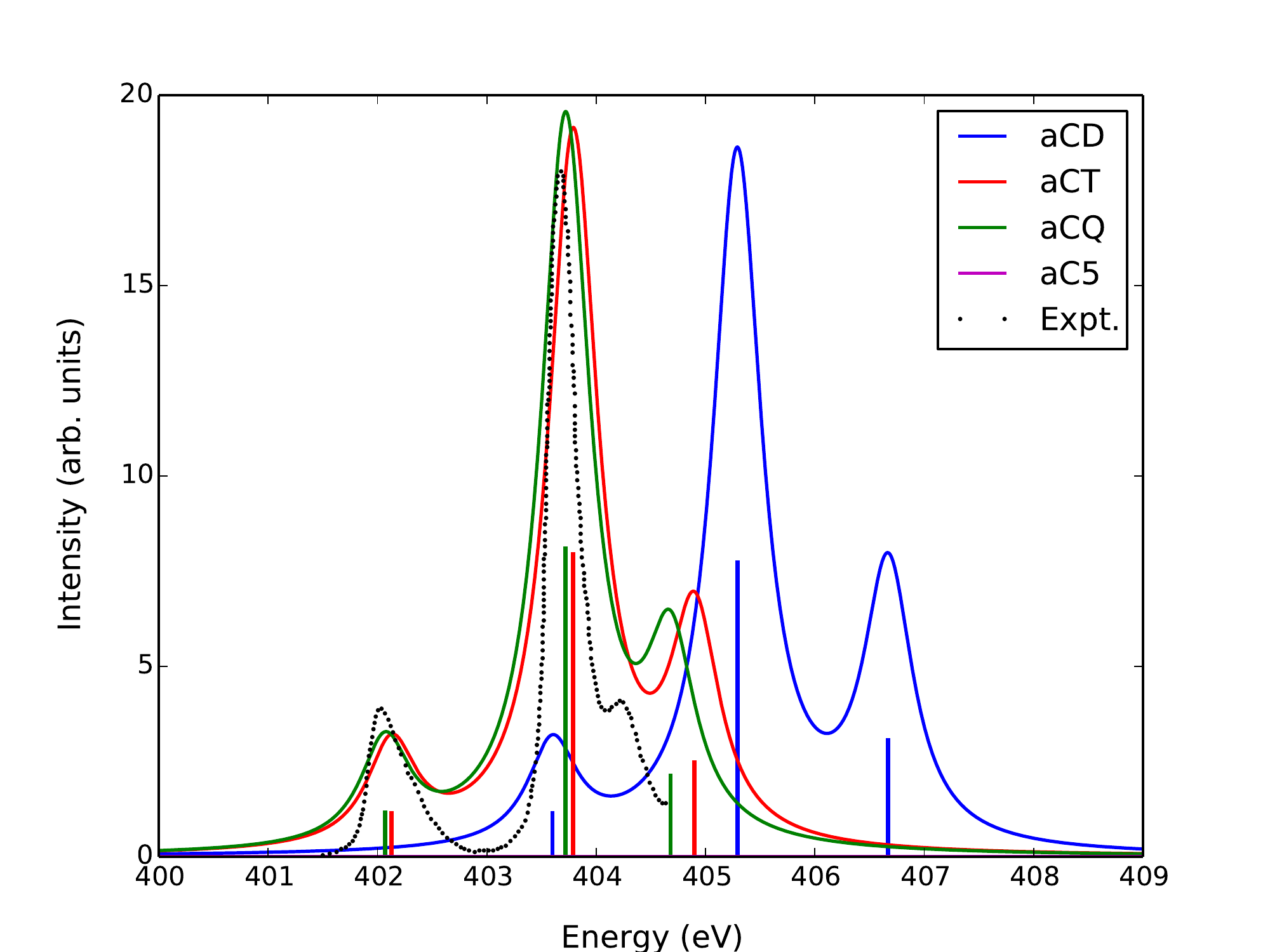}\\
\hspace*{-1.0cm}\includegraphics[scale=0.4]{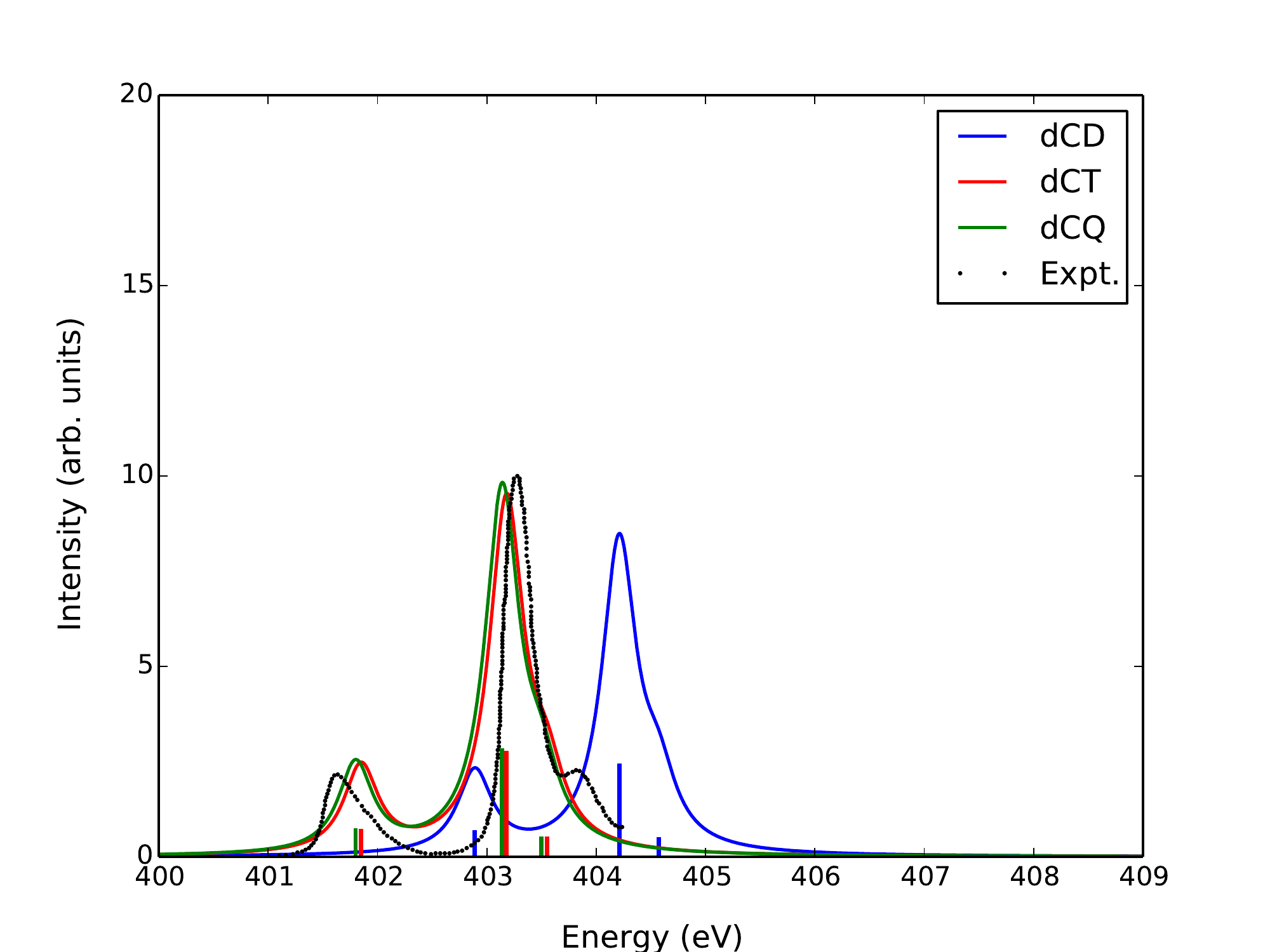}\hspace*{-0.5cm}\includegraphics[scale=0.4]{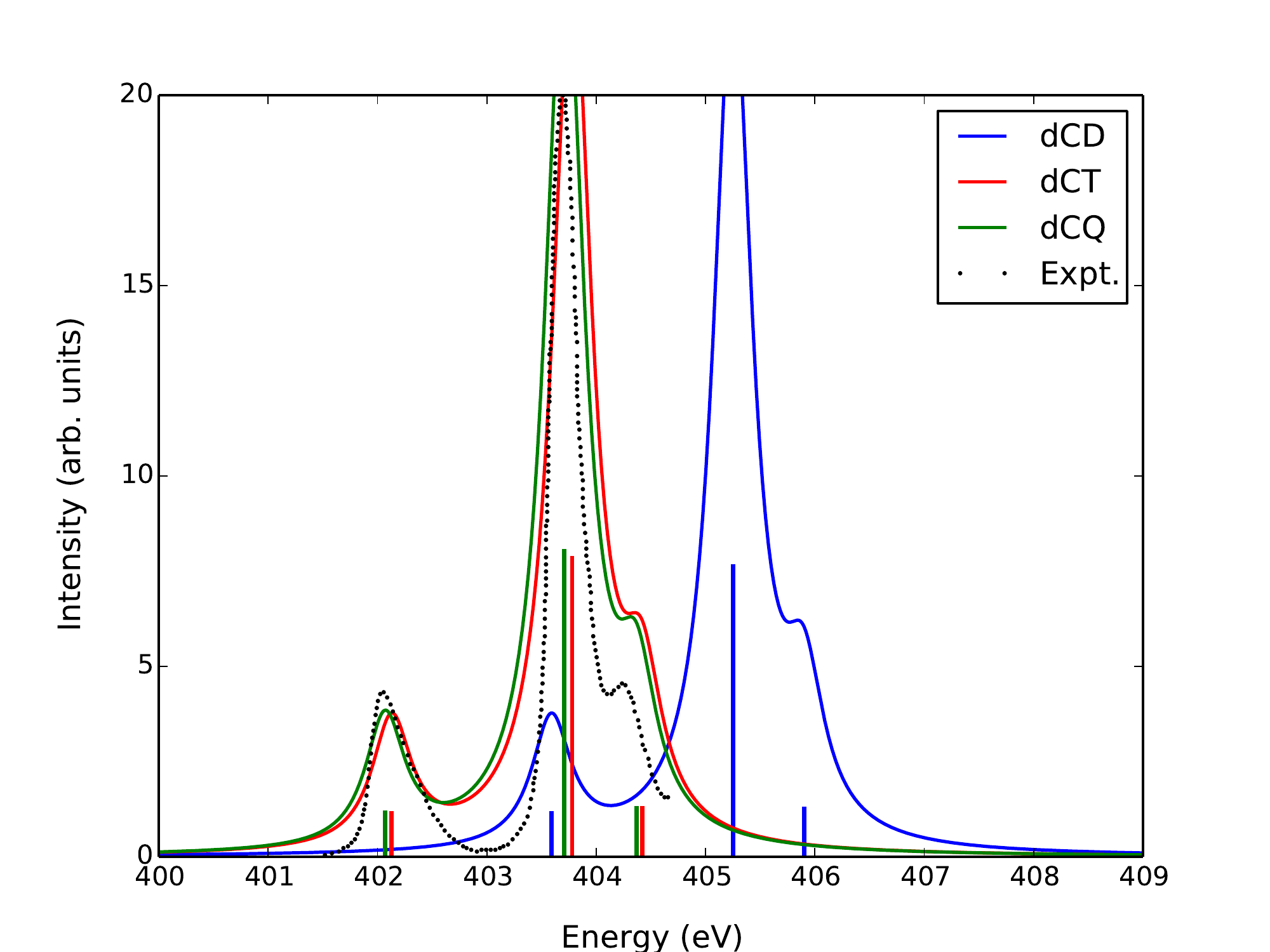}
\end{center}
\end{figure}

Spectral simulations based on oscillator strengths  for the molecules H$_2$O and
NH$_3$ computed
in the cc-pCVXZ, aug-cc-pCVXZ and d-aug-cc-pCVXZ
bases are shown in  Fig.~\ref{H2O_spectra},
and~\ref{NH3_spectra}.
They were obtained using  a Lorentzian broadening function
with a half-width-half-maximum of 0.01 a.u.

For the first two transitions of water, the CCSD oscillator strengths are  practically the same as soon as  one set of augmenting functions is added, and little affected by variation of the cardinal number, the largest differences in the spectra being
due to variations in the position of the peaks.
The situation is quite different for the third peak, which results from the combination of two excitations and has Rydberg character: many basis sets are insufficiently diffuse to yield an accurate description of the oscillator strengths, which are strongly overestimated.
Inspection of the symmetry of the excited states also reveals
that the third and fourth excited states contributing to the third spectral band
switch position energetically in the different basis sets.
An efficient strategy to have a good representation of the third band on H$_2$O
is to include Rydberg type functions, as done in Ref.~\cite{coriani2012pra}.
The CC2 oscillator strengths are more erratic, showing a relative intensity of the peaks at large variance compared to the experimental one, even in the larger basis sets.
The CC2 spectra at the O K-edge of H$_2$O are ``compressed'', due to the
underestimation   of the separation between the bands, as it can be appreciated in Fig.~\ref{H2O_spectra}. 

For carbon in CO both the CCSD and the CC2 intensity of the
first transition grows slightly within each series as the cardinal number increases.
The intensity of the second transition is roughly constant, while large variations are observed for the intensities of the third peak, in particular at CC2 level. For Oxygen,
the intense peak has similarly almost constant intensity for all bases at CCSD level, while in the
CC2 case some variations are recorded around an average value lower than the
CCSD value. The intensities of the second and third peak are extremely low for both methods,
and in the CC2 case sometimes even lower than the limit of detection.

Finally, in the case of ammonia (see Fig.~\ref{NH3_spectra}), at both levels relatively large variations in intensity are observed for the second state (the most intense transition) for the bases lacking diffuse functions, and
almost constant values for the other Dunning series.
The intensities of the first state are roughly constant
with the exception of in the smallest bases, which
yield slightly overestimated values.
Regarding the third state, the intensity decreases in the singly augmented sets
and is almost constant for the doubly augmented ones,
an indication of the greater sensitivity of this state to the presence
of diffuse functions. The best agreement with the experimental
profile is found for the d-aug-cc-pCVQZ basis set (we did not compute the aug-cc-pCV5Z and d-aug-cc-pCV5Z oscillator strengths).
At CC2 level, there is  in general a greater variation in the intensity,
which is significantly smaller than in the CCSD case.
Also at the N K-edge of NH$_3$, the peak separation is underestimated,
yielding ``squeezed''  spectral profiles, compared to both CCSD and experiment.

\subsection{Ionization energies}

Tables~\ref{CoreIPs1} and \ref{CoreIPs2} contain the results of the core ionization energies for different basis sets in the CC hierarchy up to CCSDT.

\begin{table}[htb]
\caption{Core ionization energies of water and ammonia using different standard
basis sets and the hierarchy of CC methods CC2, CCSD, CC3 and CCSDT.
Experimental values are 539.78~eV for H$_2$O 
and 405.6 eV for NH$_3$~\cite{SchirmerExp}.
\label{CoreIPs1}}
\centering
\scriptsize
\begin{tabular}{l|cccc|cccc} \hline
     &   \multicolumn{4}{c|}{H$_2$O} &
         \multicolumn{4}{c}{NH$_3$}
         \\\cline{2-9}
Basis set & CC2     & CCSD   & CC3    & CCSDT  & CC2    & CCSD   & CC3     & CCSDT \\
\hline
 VDZ        & 539.15  & 543.31 & 541.21 & 541.77 & 406.58 & 408.66 & 407.15 & 407.45\\
 VTZ        & 537.56  & 540.67 & 538.54 & 539.04 & 404.65 & 406.30 & 404.69 & 404.90\\
 VQZ        & 537.57  & 540.77 & 538.33 & 538.94 & 404.69 & 406.40 & 404.60 & --\\
 V5Z        & 537.55  & 540.85 & 538.25 &   --     & 404.70 & 406.49 & 404.59 & -- \\
 \hline
 CVDZ       & 539.15  & 542.70 & 540.46 & 540.97 & 406.07 & 408.16 & 406.50 & 406.76 \\
 CVTZ       & 537.92  & 541.14 & 538.90 & 539.40 & 405.05 & 406.82 & 405.12 & 405.33 \\
 CVQZ       & 537.91  & 541.26 & 538.76 &  --      & 405.04 & 406.89 & 405.04 & -- \\
 CV5Z       & 537.90  & 541.35 & 538.70 & --       & 405.04 & 406.95 & -- & -- \\
 \hline
 aVDZ       & 539.88  & 544.08 & 541.31 & 542.31 & 406.59 & 409.01 & 407.24 & 407.71\\
 aVTZ       & 537.61  & 541.00 & 538.49 & 539.23 & 404.70 & 406.51 & 404.72 & 405.02 \\
 aVQZ       & 537.59  & 540.91 & 538.31 & 539.02 & 404.71 & 406.49 & 404.61 & 404.89 \\
 aV5Z       & 537.56  & 540.89 & 538.24 & --       & 404.71 & 406.52 & -- & -- \\
 \hline
  aCVDZ     & 539.27  & 543.45 & 540.56 & 541.51 & 406.09 & 408.52 & 406.60 & 407.02\\
 aCVTZ      & 537.98  & 541.47 & 538.87 & 539.61 & 405.10 & 407.03 & 405.17 & 405.47\\
 aCVQZ      & 537.93  & 541.40 & 538.73 & 539.46 & 405.06 & 406.98 & 405.05 & 405.34\\
 aCV5Z      & 537.91  & 541.38 & 538.69 &   --     & 405.05 & 406.97 &    --    & --  \\
\hline
6-311G         & 538.16 & 541.07 &   539.15  &--& 405.19 & 406.77 & 405.32 & -- \\
6-311G**       & 537.99 & 540.92 &  539.01 &--& 405.02 & 406.61 & 405.15 & -- \\
6-311++G**     & 538.09 & 541.46 & 538.97 &-- & 405.09 & 406.92 & 405.19 & -- \\ \hline
\end{tabular}
\end{table}

\begin{table}[htb]
\caption{Carbon monoxide. Core ionization energies of oxygen and carbon using different standard basis sets and the hierarchy of CC methods CC2, CCSD, CC3 and CCSDT.
Experimental values are 296.2 eV for C 
and 542.5 eV for O~\cite{SchirmerExp}.
\label{CoreIPs2}}
\centering
\scriptsize
\begin{tabular}{l|cccc|cccc} \hline
     &   \multicolumn{4}{c|}{Carbon} &
         \multicolumn{4}{c}{Oxygen}
         \\\cline{2-9}
Basis set & CC2    & CCSD   & CC3   &CCSDT & CC2    & CCSD & CC3 & CCSDT\\
\hline
 VDZ        & 299.02 & 299.32 & 298.53 & 298.49  & 542.01 & 546.59 & 543.89 & 544.72\\
 VTZ        & 297.17 & 296.98 & 295.99 & 295.88  & 539.92 & 543.71 & 541.19 & 541.83\\
 VQZ        & 297.26 & 297.08 & 295.98 & 295.88  & 539.92 & 543.68 & 541.66 & 541.66\\
 V5Z        & 297.27 & 297.12 & --       &   --      & 539.89 & 543.67 &    --    & -- \\
\hline
 CVDZ       & 298.46 & 298.81 & 297.80 & 297.73 & 541.40 & 545.97 & 543.14 & 543.91  \\
 CVTZ       & 297.64 & 297.54 & 296.48 & 296.38 & 540.28 & 544.18 & 541.55 & 542.19   \\
 CVQZ       & 297.65 & 297.59 & 296.44 & 296.34 & 540.25 & 544.18 & -- & -- \\
 CV5Z       & 297.66 & 297.61 &   --     &    --    & 540.24 & 544.18 &  -- &--\\
%
%
 \hline
 aVDZ       & 299.04 & 299.34 & 298.49 & 298.46 & 542.16 & 546.83 & 543.98 & 543.93 \\
 aVTZ       & 297.20 & 297.04 & 296.00 & 295.90 & 539.96 & 543.79 & 541.18 & 541.86 \\
 aVQZ       & 297.28 & 297.11 & 295.999& 295.90 & 539.94 & 543.71 & 541.02 & 541.68 \\
 aV5Z       & 297.29 & 297.14 &  --      & --       & 539.90 & 543.69 & --&--\\
 \hline
 aCVDZ      & 298.50 & 298.84 & 297.80 & 297.76 & 541.54 & 546.20 & 543.22  & 544.12 \\
 aCVTZ      & 297.68 & 297.62 & 296.53 & 296.43 & 540.32 & 544.27 & 541.57  & 542.25 \\
 aCVQZ      & 297.66 & 297.61 & 296.44 & 296.36 & 540.27 & 544.21 & 541.44  & 542.12 \\
 aCV5Z      & 297.66 & 297.62 &  --      &   --     & 540.25 & 544.19 &   --      &  -- \\
 \hline
\end{tabular}
\end{table}

Inspection of the results clearly reveals, also for the IE, the inaccuracy of the double zeta basis sets for the core IE: for all three edges (C, N and O) and at all CC levels
the X=D basis sets overestimate the IEs by in between 1 to 3 eV. The largest improvement is observed for X=T, whereas going beyond triple-$\zeta$ has either moderate or negligible effect, and so also does the inclusion of single augmentation.

Moving along the CC hierarchy, we note that the CCSD IEs are significantly larger (1.5--4 eV) than the CC2 ones for both O and N, whereas for the C K-edge they are just slightly smaller (a few tenths of eV). Inclusion of triple excitations at the approximate CC3 lowers the IEs by $\approx$2.5--2.8 eV for O, by $\approx$1.6--1.9 eV for N, and $\approx$1.0 eV for C. Inclusion of the full treatment of triples by CCSDT, on the other hand, increases the CC3 results by a constant amount (independent of the basis set) of $\approx$0.7 eV for the O edge, $\approx$0.3 eV for the N edge; the C-edge IEs, on the other hand, are further reduced by 0.1 eV.

Comparing with the experimental results, we observe that in the case of Carbon
both CC2 and CCSD overestimate the IE;  CC3 and CCSDT either overestimate or underestimate the IE, depending on the basis set, by a few tenths of eV.

The core IEs of the two types of oxygen K-edge (H$_2$O and CO) are significantly understimated  (ca 2 eV) at the CC2 level, and overestimated (1.5--2 eV) at the CCSD level. For X$\geq$ T,
the CC3 results are 1.0-1.5 eV lower than experiment.
The CCSDT results in the largest core-valence set are about $-$0.3 eV from experiment (and relativistic effects are $+$0.3 eV).

The N-edge IE is underestimated by 0.5--1.0 eV at the CC2 level, and overestimated by 1.0--1.5 eV
at the CCSD level. The CC3 estimates are very similar to the CC2 ones, and around
$-$0.5 eV off in the core-valence bases. The agreement with experiment is further improved by full inclusion of triple excitations: in the aug-cc-pVQZ basis the CCSDT IE is $-$0.26 eV lower than the experimental IE.

The importance of scalar relativistic effects is illustrated in Table~\ref{IPs_relativity}. As previously observed~\cite{coriani2012pra,myhre2018N2,lan2019}, the relativistic effect is core-specific and practically the same independent
of the chosen method and basis set.
The effect is to increase the IEs in all cases, which can be ascribed to the contraction, and thereby stabilization, of the core orbitals.

\begin{table}[ht]
\caption{Comparison of relativistic and non-relativistic results for the IEs (eV).
\label{IPs_relativity}}
\small
\centering
\begin{tabular}{l|ccc|ccc|}
\cline{2-7}
 & Nonrel & SFX2C-1e & $\Delta$Rel
 &Nonrel&SFX2C-1e&$\Delta$Rel \\\cline{2-7}
& \multicolumn{3}{c|}{O(H$_2$O)}
& \multicolumn{3}{c|}{N(NH$_3$)}\\
 \hline
CCSD/aVTZ & 541.03 & 541.39 & 0.36
                     & 406.51 & 406.73 & 0.22\\
CCSD/aVQZ & 540.91 & 541.30 & 0.39
                      & 406.50 & 406.71 & 0.21\\
CCSD/aV5Z & 540.90 & 541.28 & 0.38
                     & 406.52 & 406.73 & 0.21\\
CCSD/aCVTZ & 541.48 & 541.86 & 0.38
                     & 407.03 & 407.24 & 0.21\\
CCSD/aCVQZ & 541.40 & 541.79 & 0.39
                     & 406.99 & 407.20 & 0.21\\
CCSD/aCV5Z & 541.38 & 541.77 & 0.39
                       & 406.98 & 407.19 & 0.21\\
\hline
CCSDT/aVTZ & 539.23 & 539.61 & 0.38
                         & 405.02 & 405.23 & 0.21
                        \\
CCSDT/aVQZ & 539.02 & 539.40 & 0.38
                       & 404.89 & 405.10 & 0.21
                        \\
CCSDT/aCVTZ & 539.61 & 540.00 & 0.39
                           & 405.47 & 405.68 & 0.21
                           \\
                       CCSDT/aCVQZ & 539.46 & 539.85 & 0.39
                          & 405.34& 405.55& 0.21
                          \\
\hline
& \multicolumn{3}{c}{O(CO)}
& \multicolumn{3}{c|}{C(CO)}\\\hline
CCSD/aVTZ 
                     & 543.79 & 544.17 & 0.38
                     & 297.04 & 297.14 & 0.10
                     \\
CCSD/aVQZ 
                      & 543.71 & 544.10 & 0.39
                      & 297.11 & 297.21 & 0.10
                      \\
CCSD/aV5Z 
                     & 543.69 & 544.07 & 0.38
                     & 297.14 & 297.24 & 0.10
                     \\
CCSD/aCVTZ 
                     & 544.27 & 544.65 & 0.38
                     & 297.62 & 297.72 & 0.10
                     \\
CCSD/aCVQZ 
                     & 544.21 & 544.59 & 0.38
                     & 297.61 & 297.71 & 0.10
                     \\
CCSD/aCV5Z 
                       & 544.19 & 544.57 & 0.38
                       & 297.62 & 297.72 & 0.10
                       \\
\hline
CCSDT/aVTZ 
                        & 541.86 & 542.25 & 0.39
                        & 295.90 & 295.99 & 0.09
                         \\
CCSDT/aVQZ 
                        & 541.68 & 542.06 & 0.38
                        & 295.90 & 295.99 & 0.09
                        \\
CCSDT/aCVTZ 
                         & 542.25 & 542.63 & 0.38
                          & 296.43 & 296.53  & 0.10
                           \\
CCSDT/aCVQZ 
                          & 542.12 & 542.50 & 0.38
                          & 296.36 & 296.46 & 0.10
                          \\
\hline
\end{tabular}
\end{table}
\section{Concluding remarks}
We have carried out a coupled cluster investigation of the performance of the
standard hierarchy CC2-CCSD-CC3-CCSDT in connection with conventional
Dunning correlation-consistent and Pople basis sets to yield accurate vertical core excitation energies (and strengths) and core-ionization energies for N, C and O K-edges in the prototypical molecules H$_2$O, CO and NH$_3$.
Complete basis set limit values 
have been derived.

The use of singly augmented triple-zeta basis sets
is sufficiently accurate for the  low-energy core
excitations, which are of limited or no Rydberg character.
The Pople set 6-311++G**
set provides results of quality almost comparable as the
aug-cc-pVTZ set, but at lower computational cost.

\section*{Acknowledgements}

S.C. acknowledges financial support from the AIAS-COFUND program
(Grant Agreement No. 609033) and from the Independent Research Fund Denmark, DFF-Forskningsprojekt2 (Grant No. 7014--00258B).
H.K. acknowledges financial support from the FP7-PEOPLE-2013-IOF
funding scheme (Project No. 625321) and a visiting professorship
grant at DTU Chemistry from the Otto M{\o}nsted Foundation.
RHM acknowledges computer time from NOTUR through Project No. nn2962k.
The COST Actions No. MP1306 ``Modern Tools
for Spectroscopy on Advanced Materials (EUSPEC)'',
CM1204 ``XUV/X-ray light and fast ions for ultrafast
chemistry (XLIC)'' and CM1405 ``MOLIM--Molecules in Motion''
are also acknowledged.
\clearpage
\clearpage

\clearpage

\clearpage
\clearpage

\clearpage
%

\clearpage

\clearpage
\bibliographystyle{achemso}
\bibliography{Tesi}
\end{document}


\title{Supplementary Information.
An analysis of the performance of 
coupled cluster methods for core excitations and core ionizations using standard basis sets}
\author{Johanna P. Carbone}
\affiliation{Dipartimento di Scienze Chimiche e Farmaceutiche, 
Universit\`a degli Studi di Trieste, I-34127, Trieste, Italy}
\author{Lan Cheng}
\affiliation{Department of Chemistry,  Krieger School of Arts and Sciences, Johns Hopkins University, Baltimore, MD 21218, USA}
\author{Rolf H. Myhre}
\affiliation{Department of Chemistry, Norwegian University of Science and Technology, N-7491 Trondheim, Norway}
\author{Devin Matthews}
\affiliation{Department of Chemistry, Southern Methodist University, Dallas, Texas 75275, United
States}
\author{Henrik Koch}
\affiliation{Department of Chemistry, Norwegian University of Science and Technology, N-7491 Trondheim, Norway}
\affiliation{Scuola Normale Superiore, I-56126, Pisa, Italy}
\author{Sonia Coriani}
\thanks{Corresponding author. Email: soco@kemi.dtu.dk}
\affiliation{Department of Chemistry, Technical University of Denmark, DK-2800 Kgs. Lyngby, Denmark} 

\begin{abstract}
Supplementary information: tables of computed excitation energies and strengths.
\end{abstract}

\maketitle

\begin{table}[hb]
\caption{Cartesian coordinates of the three molecoles (Bohr).
\label{Coordinate}}
\centering
\footnotesize
\begin{tabular}{|c|ccc|ccc|}
 \hline
     H$_{2}$O & $x$ & $y$ & $z$ \\\hline
     O       & 0.0000000000 & 0.0000000000 & $-$0.1239093563 \\  
     H$_{1}$ & 0.0000000000 & 1.4299372840 &    0.9832657567 \\\hline
 NH$_{3}$ &            &               &         \\\hline  
N       & 0.000000000 &    0.0000000000 & $-$0.1277338768 \\
H$_{1}$ & 0.000000000 & $-$1.7699234622 &    0.5915930938\\
H$_{2}$ &1.5327986810 &    0.8849617311 &    0.5915930938\\\hline
     CO &           &           &         \\\hline  
     O  & 0.000000  & 0.000000  &    0.913973 \\
     C  & 0.000000  & 0.000000  & $-$1.218243\\\hline
\end{tabular}
\end{table}

\begingroup
\begin{table}
\caption{H$_2$O, Oxygen K-edge. CC results for the first four core excitations
using different families of standard basis sets. 
Dunning's cc-pVXZ sets are specified as V$X$Z;
the cc-pCVXZ ones as  CV$X$Z; the corresponding augmented 
and double augmented ones as aV$X$Z, aCV$X$Z, dV$X$Z, dCV$X$Z.
In the case of Pople's basis sets,  G stands for 6-311G,
G** for 6-311G** and ++G** stands for 6-311++G**. 
Experimental values: $\omega_1=534.0$;
$\omega_2=535.9$ and
$\omega_3=537.1$~eV.\label{H2O_energies}}
\begin{adjustbox}{angle=-90}
\scriptsize
\centering
\begin{tabular}{|l|cccc|cccc|cccc|cccc|} \hline
     & \multicolumn{4}{c|}{CC2} 
     & \multicolumn{4}{|c|}{CCSD} 
     & \multicolumn{4}{|c|}{CC3}
     & \multicolumn{4}{|c|}{CCSDT}\\\cline{2-17}
Basis & $\omega_1$ & $\omega_2$ & $\omega_3$ & $\omega_4$
      & $\omega_1$ & $\omega_2$ & $\omega_3$ & $\omega_4$
      & $\omega_1$ & $\omega_2$ & $\omega_3$ & $\omega_4$
      & $\omega_1$ & $\omega_2$ & $\omega_3$ & $\omega_4$
      \\\hline
VDZ     & 537.61 & 539.61 & 551.70$^a$  & 552.97$^b$  
      & 538.37 & 540.19 & 552.23$^a$  & 553.28$^b$
      & 537.13 & 539.08 & 550.79 & 551.91 
      & 537.29 & 539.23 & 550.94 & 552.06
        \\  
VTZ      & 534.74 & 536.69 & 544.10$^a$ & 545.30$^b$
        & 535.37 & 537.18 & 544.62$^a$ & 545.70$^b$
        & 533.97 & 535.90 & 543.08      & 544.17
        & 534.10 & 536.03 & 543.20 & 544.30
        \\      
VQZ       & 534.48 & 536.37 & 542.32(b2) & 543.18(a1)
        & 535.25 & 537.05 & 543.08$^a$ & 543.53$^b$
        & 533.59 & 535.49 & 541.26      & 542.04
        & 533.77 & 535.67 & 541.46 & 542.41
        \\
         
V5Z      & 534.20 & 536.01 & 540.26(a1)& 540.57(b2)
        & 535.17 & 536.97 & 541.24$^b$ & 541.61$^a$
        & 533.34 & 535.20 & 539.46 & 539.64
         & -      & -      &  - & -
        \\
        \hline
aVDZ          & 536.35 & 537.91 & 539.94$^b$ & 540.50$^c$
            & 538.25 & 540.05 & 542.20$^b$ & 542.34$^c$
            & 536.34 & 538.12 & 540.16 & 540.57	
            & 536.84     & 	538.68     &  540.75 & 541.00
            \\
aVTZ          & 533.96 & 535.51 & 537.02$^b$ & 537.49$^c$
            & 535.22 & 537.01 & 538.81$^b$ & 538.91$^c$
            & 533.43 & 535.20 & 536.84 & 537.13
            & 533.75	 & 535.57	& 537.27	& 537.44
            \\
aVQZ        & 533.92 & 535.47 & 536.73$^b$ & 537.07$^c$
            & 535.11 & 536.89 & 538.54 & 538.57
            & 533.23 & 534.99 & 536.44 & 536.61
            & 533.52 & 535.34 & 536.86 & 536.95
            \\
aV5Z          & 533.88 & 535.42 & 536.43$^b$ & 536.63$^c$
            & 535.09 & 536.86 & 538.33$^a$ & 538.34$^b$
            & 533.16 & 534.92 & 536.15 & 536.23
            & -      & -      &  - & -
            \\ 
\hline
dVDZ            & 536.32 & 537.84 & 538.34(b1) & 538.37(a1)
              & 538.22 & 540.02 & 541.20$^c$ & 541.30$^b$
              & 536.31 & 538.08 & 538.90 & 538.95
              & -      & -      &  - & -
              \\
dVTZ            & 533.95 & 535.48 & 535.99$^c$ & 536.02$^b$
              & 535.22 & 537.00 & 538.14 & 538.23
              & 533.41 & 535.17 & 536.00 & 536.06
              & -      & -      &  - & -
              \\
dVQZ            & 533.91 & 535.45 & 535.96$^c$ & 535.97$^b$
              & 535.11 & 536.89 & 538.04(b1) & 538.14(a1)
              & 533.22 & 534.98 & 535.82 & 535.87 
              & -      & -      &  - & -\\
dV5Z            & 533.87 & 535.41 & 535.91$^c$ & 535.95$^b$
              & 535.09 & 536.86 & 538.02$^b$ & 538.12$^b$
              & 533.16 & 534.92 & 535.75 & 535.81 
              & -      & -      &  - & -\\\hline
CVDZ       & 536.99 & 538.99 & 551.08$^a$  & 552.33$^b$ 
         & 537.77 & 539.58 & 551.61 & 552.65$^b$ 
         & 536.39 & 538.33 & 550.03 & 551.14 
         & 536.51 & 538.46 & 550.15 & 551.26 \\
CVTZ       & 535.11 & 537.07 & 544.48$^a$ & 545.68$^b$
         & 535.81 & 537.63 & 545.08 & 546.15
         & 534.31 & 536.25 & 543.44 & 544.52 
         & 534.44 & 536.38 & 543.56 & 544.65 \\
CVQZ       & 534.84 & 536.73 & 542.69 & 543.54
         & 535.73 & 537.53 & 543.57 & 544.01
         & 534.01 & 535.91 & 541.70 & 542.47 
         & 534.20 & 536.11 & 541.90 & 542.82 \\
CV5Z       & 534.58 & 536.38&  540.63 (a1) & 540.95 (b2)
         & 535.66 & 537.45 & 541.73 & 542.09
         & 535.59 & 537.37& 539.40 & 540.10 
         & -      & -      &  - & -\\\hline
aCVDZ          & 535.74 & 537.29 & 539.32(a1) & 539.88(b1)
             & 537.63 & 539.44 & 541.57 & 541.71
             & 535.59 & 537.37 & 539.40 & 539.81
             & 536.05 &	537.89	& 539.95	 & 540.21\\
aCVTZ          & 534.35 & 535.90 & 537.40$^b$ &  537.87
             & 535.69 & 537.47 & 539.28 & 539.37
             & 533.81 & 535.58 & 537.23 & 537.51 
             & 534.12 &	535.95	& 537.65 & 	537.82\\
aCVQZ     & 534.27 & 535.82 & 537.09 & 537.42
         & 535.59 & 537.37 & 539.03(a1) & 539.06(b1)
         & 533.65 & 535.42 & 536.87 & 537.04 
         & 533.96 & 535.78 & 537.30 & 537.39
             \\
aCV5Z      & 534.24 & 535.79 & 536.79(a1)& 536.99(b1)
             & 535.57 & 537.34 & 538.81(b1) & 538.83(a1)
             & 533.60 & 535.37 & 536.60 & 536.68 
             & -      & -      &  - & -\\\hline
dCD        & 535.71 & 537.22 & 537.72(b1) & 537.76 (a1)
               & 537.60 & 539.49 & 540.57 & 540.67 (a1)
               & 535.56 & 537.33 & 538.15 & 538.20 
               & -      & -      &  - & -\\
dCT            & 534.33 & 535.86 & 536.37(b1)& 536.40(a1) 
               & 535.68 & 537.46 & 538.61(b1)& 538.71(a1)
               & 533.80 & 535.56 & 536.39 & 536.45
               & -      & -      &  - & -\\
dCQ        & 534.27 & 535.80 & 536.30(b1) &536.32(a1)
           & 535.59 & 537.37 & 538.53(b1)& 538.63(a1)
           & 533.65 & 535.41 & 536.25 & 536.30 
           & -      & -      &  - & -\\
dC5        & 534.24 & 535.77 & 536.27(b1) & 536.31(a1)
           & 535.57 & 537.34 & 538.51(b1) & 538.61(a1)
           & 533.60 & 535.36 & 536.20     & 536.26
           & -      & -      &  - & -
               \\\hline
G  (S)  
     & 535.34 & 537.47 & 545.09 & 546.11(a1)
     & 535.68 & 537.65 & 545.27(b2) & 546.19(a1)
     & 534.51 & 536.50 & 543.80 & 544.86
     & 534.57 & 536.67 & 544.01 & 544.91
     \\
G** (S)
    & 535.29 & 537.30 & 544.84 &546.01
    & 535.72 & 537.58 & 545.11 &  546.18
    & 534.49 & 536.60 & 543.97 & 544.87
    & 534.62 & 536.60 & 543.89 & 544.95
    \\
G** (C)
    & 534.79 & 536.83 & 544.35(b2) & 545.52(a1)
    & 535.19 & 537.08 & 544.57(b2) & 545.66(a1)
    & 534.01 & 536.02 & 543.30 & 544.37
    & 534.09 & 536.09 & 543.36 & 544.43
    \\
$^{++}$G**(S)
      & 534.52 & 536.12 & 538.94(a1)&539.30(b2)
      & 535.70 & 537.49 & 540.21(a1)&540.40(b1)
      & 533.99 & 535.76 & 538.45 & 538.71
      & 534.33 & 536.15 & 538.84(a1) & 539.10(b2)
\\
$^{++}$G**(C)
      & 534.05 & 535.66 & 538.46 & 538.83
      & 535.18 & 536.98 & 539.69 & 539.89
      & 533.51 & 535.29 & 537.97(a1) & 538.23(b2)
      & 533.81 & 535.64 & 538.32(a1) & 538.58(b2)
      \\\hline
\end{tabular}
\end{adjustbox}
\end{table}
\endgroup

\begin{table}[h]
\caption{H$_2$O, Oxygen K-edge. CC2 and CCSD results for the oscillator strengths of 
the first four core excitations  using different families of standard basis sets.
In the case of Pople's basis sets,  (S) stands for spherical and (C) stands for cartesian. 
\label{H2O_forze}}
\centering
\scriptsize
\begin{tabular}{|l|cccc|cccc|} \hline
     & \multicolumn{4}{c|}{CC2} &
      \multicolumn{4}{|c|}{CCSD} \\\cline{2-5}\cline{6-9}
Basis & $f_1$ & $f_2$ & $f_3$ & $f_4$
          & $f_1$ & $f_2$ & $f_3$ & $f_4$ \\\hline
cc-pVDZ     & 0.0124 & 0.0305 & 0.0196(b2) & 0.030 (a1)
       &0.0155 & 0.0372&0.0150(b2) & 0.0317(a1)\\
cc-pVTZ     & 0.0110 & 0.0278 & 0.0110(b2) & 0.0158(a1)
       &0.0144 & 0.0358&0.0072(b2) & 0.0197(a1)\\
cc-pVQZ     & 0.0096 & 0.0230 & 0.0094 & 0.0164(a1)
        &0.0136 & 0.0328&0.0065(b2) & 0.0364(a1)\\
cc-pV5Z     & 0.0083 & 0.0183 & 0.0065(a1) &0.0076(b2)
       &0.0129 & 0.0300&0.0173(a1) & 0.0304(b2)
\\\hline
aug-cc-pVDZ    & 0.0060 & 0.0085& 0.0122 & 0.0127
        &0.0124 & 0.0256&0.0141 &  0.0171\\
aug-cc-pVTZ   & 0.0063 & 0.0092& 0.0098 & 0.01097
        &0.0120 & 0.0254&0.0117 & 0.0150 \\
aug-cc-pVQZ   & 0.0064 & 0.0092& 0.0080 & 0.0090(b1)
        &0.0122 & 0.0257&0.0097 & 0.0131\\
aug-cc-pV5Z    & 0.0065 & 0.0090& 0.0054(a1)& 0.0069 (b1) 
        &0.0123 & 0.0259&0.0110(b1)& 0.00751(a1)
\\\hline
d-aug-cc-pVDZ & 0.0060 & 0.0073& 0.0029&0.0014
      & 0.0123 & 0.0255&0.0079 & 0.0046 \\
d-aug-cc-pVTZ & 0.0064 & 0.0085& 0.0030 & 0.0015(a1)
     &0.0121 & 0.0253&0.0076 & 0.00472 \\
d-aug-cc-pVQZ & 0.0065 & 0.0087& 0.0030 & 0.00137(a1) 
      &0.0122 & 0.0257&0.0076 & 0.0047(a1)\\ 
d-aug-cc-pV5Z & 0.0065 & 0.0088& 0.0029& 0.00154(a1)
&0.0123 & 0.0258&0.0076(b1) & 0.0047(a1)
\\\hline
cc-pCVDZ &0.0122 & 0.0303&0.0195 (b2)& 0.030 (a1)
   &0.0154 & 0.0371&0.0150(b2) & 0.0316(a1)\\
cc-pCVTZ &0.0110 & 0.0276&0.0110& 0.0155
   &0.0145 & 0.0358&0.0073& 0.0197\\
cc-pCVQZ &0.0095 & 0.0229&0.0094& 0.0137
&0.0136 & 0.0329&0.0065 & 0.0365 \\
cc-pCV5Z &0.0083 &  0.0182&0.0064 (a1) & 0.0076 (b2)
   &0.0130 & 0.0301&0.0173(a1) & 0.0305 (b1)
\\\hline
aug-cc-pCVDZ & 0.0060 &0.0084&0.0121 & 0.0127
        &0.0123 & 0.0254&0.0141& 0.0170 \\
aug-cc-pCVTZ & 0.0063 &0.0091&0.00976& 0.0108
        &0.0121 & 0.0256&0.0116 & 0.0150\\
aug-cc-pCVQZ & 0.0064 &0.0091&0.0079 & 0.0089  
    &0.0123 & 0.0259&0.0097 & 0.0131 \\
aug-cc-pCV5Z & 0.0065 &0.0089&0.0054(a1)&0.0069(b1)
    &0.0123 & 0.0260&0.0111(b1)&0.0075(a1)
\\\hline
d-aug-cc-pCVDZ &0.0059 & 0.0073 &0.0029(b1) & 0.00143(a1)
    &0.0123 & 0.0254 &0.0078 & 0.0046(a1)\\
d-aug-cc-pCVTZ &0.0064 & 0.0084 &0.00295(b1)& 0.00149(a1)
    &0.0122 & 0.0255 &0.00762(b1)& 0.00474(a1)\\
d-aug-cc-pCVQZ &0.0065 & 0.0086 & 0.0030(b1) & 0.0013(a1)
        &0.0123    & 0.0258&0.0077 & 0.0047\\
d-aug-cc-pCV5Z &0.0065 & 0.0086 &0.0029(b1) & 0.0015(a1)
    &0.0123 & 0.0260 &0.0077(b1) & 0.0047(a1)
\\\hline
6-311G   &0.0125 & 0.0302&0.0169& 0.0175
    &0.0160 & 0.0385&0.0111& 0.0168 \\
6-311G**(S) 
    &0.0112 &0.0288& 0.0145 & 0.01569
    &0.0142 & 0.0361&0.0096 & 0.0153 \\
6-311G**(C) 
    &0.0116 & 0.0296 & 0.0154 & 0.0143 
    &0.0146 & 0.0366 & 0.0095 & 0.0150 \\
6-311$^{++}$G**(S) 
      &0.0066&0.0111&0.0159 & 0.0152
      &0.0121&0.0259&0.0177 & 0.0196 \\
6-311$^{++}$G**(C)
      &0.0066&0.0111&0.0162&0.0157
      &0.0121&0.0255&0.0181&0.01996 \\
\hline
\end{tabular}
\end{table}

	
\begin{table}[h]
{\caption{CO, Carbon K-edge. CC results for the first three vertical core excitations using different families of standard basis sets.
Experimental values: 287.4~eV, 292.5~eV and 293.4~eV.
Dunning's cc-pVXZ sets are specified as VXZ; the augmented families are  
indicated as aVXZ and dVXZ, and the core-polarized families as CVXZ, aCVXZ and dCVXZ.
In the case of Pople's basis sets,  G stands for 6-311G,
G** for 6-311G** and ++G** stands for 6-311++G**. 
\label{CO_C_energies}}}
\centering
\scriptsize
\begin{tabular}{|l|ccc|ccc|ccc|ccc|} \hline
     &   \multicolumn{3}{c|}{CC2}   &
         \multicolumn{3}{|c|}{CCSD} & 
         \multicolumn{3}{|c|}{CC3} & \multicolumn{3}{|c|}{CCSDT}\\\cline{2-13}
Basis set&   
$\omega_1$ &$\omega_2$ &	 $\omega_3$ &	 
$\omega_1$ 	& $\omega_2$ &	 $\omega_3$ 	&
$\omega_1$ 	& $\omega_2$ &	 $\omega_3$ &
$\omega_1$ 	& $\omega_2$ &	 $\omega_3$ \\
\hline
VDZ 	& 291.31 & 301.54 & 305.67	
        & 290.27 & 300.49 & 302.32 
        & 290.11 & 299.64** &  301.71** 
        & 290.03 & 298.42** &  298.95** \\
VTZ	& 289.05 & 297.52 & 300.17	
        & 287.68 & 296.13 & 298.80 
        & 287.34 & 295.33 & 297.28
        & 287.21 & 295.11 & 295.78 \\
VQZ 	& 289.01 & 296.54 & 298.52	
        & 287.60 & 295.23 & 297.24 
        & 287.20 & 294.28 & 296.33 
        & -      & -      &  - \\
V5Z	& 288.96 & 295.75 & 296.93	
        & 287.57 & 294.53 & 295.79 
        & 287.14 & 293.52 & 294.83
    & -      & -      &  - \\\hline
aVDZ &  291.10& 296.51& 297.5	
   &  290.03& 295.82& 296.93	
   &  289.88& 295.11& 296.22
   &  289.79& 294.98& 296.09 \\
aVTZ          &	289.01	& 294.55 &295.46	
            &	287.64	& 293.46 &294.50	
            &   287.29  & 292.54 &293.58 
            &   287.16  & 292.34 &293.39\\
aVQZ 	    &	289.00	& 294.55	&	295.44	
            &	287.60	& 293.48	&	294.50	
            &   287.17  & 292.48    &   293.50
            &   287.05  & 292.28 &  293.30 \\
aV5Z 	&	288.97	& 294.50	&	295.36	
            &	287.58	& 293.46	&	294.46	
            &   287.14  & 292.43  &   293.42
            & -      & -      &  -\\\hline
dVDZ          &291.05 &296.17& 296.98
	    &289.96 &295.54& 296.52
            &289.81 &294.82& 295.78
	    &298.73 &294.70& 295.67
            \\
dVTZ &289.01&294.34 &295.16
   &287.64&293.32 &294.31
   &287.28&292.39 &293.35
   & -      & -      &  -
            \\
dVQZ
&288.99	& 294.41&295.23
&287.60	& 293.39&294.37
&287.19 & 292.38&293.35
			& -      & -      &  -
            \\
dV5Z 		
&	288.97	&	294.42	&	295.25
&	287.59	&	293.42	&	294.40
& 287.15 & 292.38 & 293.35
& -      & -      &  -
            \\\hline
CVDZ 		&	290.84	&	300.97	&	305.16
			&	289.77	&	299.94	&	303.00
             & 289.49 & 299.24 & 299.28
			& 289.38 & 297.89 & 298.51
            \\
CVTZ 		&	289.54	&	297.97	&	300.65
			&	288.19	&	296.66	&	299.35
			& 287.81 & 295.79& 297.79
			& 287.68 & 296.28 & 296.84
            \\
CVQZ 		&	289.41	&	296.93	&	298.91
			&	288.06	&	295.72	&	297.73
			& 287.63 & 294.73&  296.77
			& -      & -      &  -
            \\
CV5Z 		&	289.37	&	296.14	&	297.32
			&	288.04	&	295.02	&	296.28
             & 287.59 & 293.97 & 295.29
		    & -      & -      &  -
            \\\hline
aCVDZ	&	290.64	&	295.98	&	296.97	
                &289.55	&	295.32	&	296.43	
                & 289.27 & 294.43 & 295.54
                & 289.17	& 294.27	& 295.39
                \\
aCVTZ  
& 289.52	& 295.04	&	295.96	
& 288.18	& 294.04	&	295.08	
& 287.80 & 293.06& 294.11
& 287.66	& 292.87	& 293.92
                 \\
aCVQZ    & 289.4	& 294.95& 295.84
                & 288.06& 293.97 & 295	
                & 287.62& 292.94 & 293.95
                & 287.49 & 292.74 & 293.77
              \\
aCV5Z    &289.37&294.89&295.75	
      &288.04&293.94&294.94	
      &287.58&292.88&293.87 
      & -    & -    &  -\\\hline
dCVDZ & 290.58 & 295.63 &	296.45
	& 289.48 & 295.04&	296.02	
	& 289.21 & 294.14 & 295.10
& -      & -      &  -\\
dCVTZ & 289.51	&	294.83	&	295.65	
&	288.18	&	293.9	&	294.88	
& 287.79 & 292.91& 293.88
& -      & -      &  -\\
dCVQZ & 289.4	&	294.81	&	295.63	
&	288.06	&	293.89	&	294.87	
& 287.62 &  292.84 & 293.81
& -      & -      &  -\\
dCV5Z & 289.37	&	294.81	&	295.63	&	288.03	
      &	293.9	&	294.88	
& 287.585 & 292.83 & 293.797
& -      & -      &  -\\\hline
G 	&  290.30	&  300.15	&	303.58
	&	288.86	&	298.52	&	300.97	
	&  288.70&  298.02& 298.14 
	& 288.58 & 296.90 & 297.72 \\
G** 	(S)
&289.68	&299.43	&	302.95	
&	288.31	&	297.92	&	297.95	
& 288.05&  297.30 & 297.88
& 287.93 & 296.42 & 297.08
\\
G** (C)	
& 289.36 & 299.04 &	302.61
& 287.95 & 297.45 &	300.94 	
& 287.70 & 296.84 & 297.61 
& 287.57 & 296.09 & 296.62
\\
++G** (S) & 289.66 &295.14	&	296.24	
           & 288.30 &294.00	&	295.22	
           & 288.04 & 293.17 & 294.39 
           & 287.91 & 292.97 & 294.20 \\
 ++G** (C) & 289.35 & 294.81 & 295.92
           & 287.95 & 293.61 & 294.82 	
           & 287.68 & 292.79 & 294.01  
           & 287.55 & 292.58 & 293.81 \\\hline
\end{tabular}
\end{table}

\begin{table}[htb]
\caption{CO, Carbon K-edge. CC2 and CCSD oscillator strengths for the first three core excitations using different families of standard basis sets.
\label{CO_forze_Carbon}}
\centering
\scriptsize
\begin{tabular}{|l|ccc|ccc|} \hline
     &   \multicolumn{3}{c|}{CC2} &
         \multicolumn{3}{|c|}{CCSD} \\\cline{2-4}\cline{5-7}
Basis set & $f_1$ & $f_2$ & $f_3$ 
     & $f_1$ & $f_2$ & $f_3$ \\\hline
cc-pVDZ&	0.1565	&	0.0058	&	0.1383	&	0.1468	&	0.0059	&	0.0083	\\
cc-pVTZ&	0.1644	&	0.0046	&	0.0708	&	0.1513	&	0.0043	&	0.0723	\\
cc-pVQZ&	0.1658	&	0.005	&	0.0499	&	0.1531	&	0.0049	&	0.0511	\\
cc-pV5Z&	0.1659	&	0.0054	&	0.0294	&	0.1536	&	0.0054	&	0.0308	\\
\hline
aug-cc-pVDZ&	0.1564	&	0.0044	&	0.0137	&	0.1465	&	0.0046	&	0.0152	\\
aug-cc-pVTZ&	0.1644	&	0.0043	&	0.0133	&	0.1517	&	0.0044	&	0.0151	\\
aug-cc-pVQZ&	0.1654	&	0.0041	&	0.0113	&	0.153	&	0.0043	&	0.0132	\\
aug-cc-pV5Z&	0.1658	&	0.0039	&	0.0093	&	0.1536	&	0.0042	&	0.0114	\\
\hline
d-aug-cc-pVDZ&	0.1565	&	0.0033	&	0.0061	&	0.1463	&	0.0037	&	0.0089	\\
d-aug-cc-pVTZ&	0.1644	&	0.0035	&	0.0069	&	0.1517	&	0.0038	&	0.0097	\\
d-aug-cc-pVQZ&	0.1654	&	0.0035	&	0.0068	&	0.153	&	0.0039	&	0.0096	\\
d-aug-cc-pV5Z&	0.1658	&	0.0035	&	0.0068	&	0.1536	&	0.0039	&	0.0096	\\
\hline
cc-pCVDZ&	0.154	&	0.0055	&	0.1372	&	0.1456	&	0.0056	&	0.0258	\\
cc-pCVTZ&	0.1645	&	0.0047	&	0.0706	&	0.152	&	0.0044	&	0.0718	\\
cc-pCVQZ&	0.1665	&	0.005	&	0.0497	&	0.1545	&	0.005	&	0.0508	\\
cc-pCV5Z&	0.1667	&	0.0054	&	0.0292	&	0.155	&	0.0054	&	0.0305	\\
\hline
aug-cc-pCVDZ&	0.154	&	0.0042	&	0.0138	&	0.1454	&	0.0046	&	0.0151	\\
aug-cc-pCVTZ&	0.1643	&	0.0043	&	0.0131	&	0.1521	&	0.0045	&	0.0148	\\
aug-cc-pCVQZ&	0.1662	&	0.0041	&	0.0111	&	0.1544	&	0.0043	&	0.0129	\\
aug-cc-pCV5Z&	0.1666	&	0.0039	&	0.0093	&	0.155	&	0.0042	&	0.0113	\\
\hline
d-aug-cc-pCVDZ&	0.1541	&	0.0032	&	0.0062	&	0.1452	&	0.0037	&	0.0089	\\
d-aug-cc-pCVTZ&	0.1643	&	0.0035	&	0.0067	&	0.1521	&	0.0038	&	0.0096	\\
d-aug-cc-pCVQZ&	0.1661	&	0.0035	&	0.0067	&	0.1544	&	0.0039	&	0.0096	\\
d-aug-cc-pCV5Z&	0.1666	&	0.0035	&	0.0067	&	0.155	&	0.0039	&	0.0096	\\
\hline
6-311G	&0.1738	&	0.0073	&	0.1066	&	0.1576	&	0.0067	&	0.0267	\\
6-311G**	&0.1658	&	0.006	&	0.1006	&	0.1512	&	0.0054	&	0.0812	\\
6-311++G**	&0.1658	&	0.0047	&	0.0159	&	0.1514	&	0.0046	&	0.0178	\\\hline
\end{tabular}
\end{table}

\begin{table}[htb]
\caption{CO, Oxygen K-edge. CC results for the first three core excitations  
using different families of standard basis sets.
Experimental values: 534.1, 538.8 and 539.8~eV.
Dunning's cc-pVXZ sets are specified as VXZ; the augmented 
families as aVXZ and dVXZ, and the core-polarized families as 
CVXZ, aCVXZ and dCVXZ.
In the case of Pople's basis sets, G stands for 6-311G,
G** for 6-311G** and ++G** stands for 6-311++G**. 
\label{CO_O_energies}}
\centering
\scriptsize
\begin{tabular}{l|ccc|ccc|ccc|ccc} \hline
Basis set	     
& \multicolumn{3}{c|}{CC2} 
& \multicolumn{3}{|c|}{CCSD}
& \multicolumn{3}{|c|}{CC3}
& \multicolumn{3}{|c}{CCSDT}
\\\cline{2-13}
&$\omega_1$ 	&$\omega_2$ 	&$\omega_3$ 	
&$\omega_1$ &$\omega_2$ &$\omega_3$
&$\omega_1$ &$\omega_2$ &$\omega_3$ 
&$\omega_1$ &$\omega_2$ &$\omega_3$ 
\\\hline
VDZ 	
& 537.68 & 545.71 & 551.49 
& 538.43 & 547.64 & 554.26 
& 536.83 & 545.51 & 547.17
& -- & -- & -- \\
VTZ	
& 535.08 & 541.34 & 544.66 
& 535.40 & 542.81 & 546.74 
& 533.75 & 540.93 & 544.03
     & -- & -- & -- \\
VQZ 	
& 534.95 & 540.21 & 538.49 
& 535.23 & 541.74 & 544.67 
& 533.47 & 539.76 & 541.97
& -- & -- & -- \\  
V5Z 	
    & 534.85 &	539.3 & 540.52 
    & 535.18 & 540.99 & 542.82 & 533.36 & 538.91 & 540.17
& -- & -- & -- \\\hline
aVDZ & 537.53 & 540.35 & 541.33 
   & 538.43 & 543.30 & 544.68 
   & 536.76 & 540.95 & 541.97 
   & 537.03 & 541.56 & 542.73\\  
aVTZ & 535.02	&	537.94	& 538.83 
   & 535.40 & 540.18 & 541.44 
   & 533.71 & 537.99 & 539.00
   & 533.85 & 538.40 & 539.49        \\
aVQZ & 534.92 & 537.82 & 538.66 
   & 535.24 & 540.05 & 541.26 
   & 533.46 & 537.76 & 538.74
   & 533.58 & 538.16 & 539.21\\  
aV5Z & 534.84 & 537.69 & 538.49 
   & 535.19 & 539.99 & 541.14 
   & 533.36 & 537.63 & 538.59 
    & -- & -- & --        \\\hline  
dVDZ &	537.53	& 539.88	& 540.61
   & 538.41 & 543.08 & 544.18 & 536.75 & 540.65 & 541.57
& -- & -- & -- \\ 
dVTZ &	535.01	& 537.61	& 538.35 
   & 535.39 & 540.04 & 541.14 & 533.70 & 537.79 & 538.73
& -- & -- & -- \\
dVQZ &	534.92	& 537.59	& 538.32 
& 535.24 & 539.97 & 541.06 & 533.46 & 537.63 & 538.57
& -- & -- & -- \\
dV5Z &	534.84	& 537.55	& 538.29 
& 535.19 & 539.94 & 541.03 & 533.36 & 537.56 & 538.50
& -- & -- & -- \\\hline
CVDZ &	537.05	&	545.05	& 550.82 & 537.83 & 547.00 & 553.62 & 536.08 & 544.74 & 546.54\\  
CVTZ &	535.44	&	541.67	& 545.00 & 535.83	& 543.24 & 547.19 & 534.10 & 541.26 & 543.96\\  
CVQZ &	535.29	&	540.56	& 542.86 & 535.71	& 542.23 & 545.15 & 533.88 & 540.18 & 542.24\\  
CV5Z &	535.23	&	539.66	& 540.88 & 535.68 & 541.49 & 543.32 & 533.81 & 539.36 & 540.62\\\hline 
aCVDZ & 536.90 & 539.71 & 540.70 
    & 537.82 & 542.66 & 544.05 
    & 536.00 & 540.17 & 541.21 
    & 536.25 & 540.75 & 541.92 \\ 
aCVTZ & 535.39 & 538.31 & 539.20 
    & 535.84 & 540.65 & 541.92 
    & 534.08 & 538.37 & 539.39
    & 534.21 & 538.78 & 539.88 \\  
aCVQZ & 535.27 & 538.17 & 539.00 
    & 535.71 & 540.54 & 541.75 
    & 533.87 & 538.18 & 539.16 
    & 534.01 & 538.60 & 539.21 \\  
aCV5Z & 535.22 & 538.06 & 538.85 
    & 535.68 & 540.49 & 541.64 
    & 533.81 & 538.09 & 539.04
    & -- & -- & -- \\\hline 
dCVDZ & 536.89	&	539.26	& 539.98 & 537.81	& 542.44 & 543.55 & 535.99 & 539.88 & 540.81
& -- & -- & -- \\ 
dCVTZ & 535.39	&	537.99	& 538.72 & 535.84 & 540.52 & 541.62 & 534.07 & 538.18 & 539.12
& -- & -- & -- \\  
dCVQZ & 535.26	&	537.93	& 538.67 & 535.71 & 540.46 & 541.55 & 533.87 & 538.06 & 538.99
& -- & -- & -- \\ 
dCV5Z & 535.22	&	537.91	& 538.65 & 535.68 & 540.44 & 541.54 & 533.81 & 538.02 & 538.96
& -- & -- & -- \\\hline  
G     &	535.95 & 543.55	& 547.91 
      & 536.13 & 545.11 & 550.19 
      & 534.71 & 543.05 & 544.02
      & 534.78 & 542.54 & 547.65 \\  
G** (S) & 535.77 & 543.31 & 547.88 
        & 535.99 & 544.70 & 549.86 
        & 534.47 & 542.76 & 543.96
        & 534.59 & 542.60 & 547.53 \\  
G** (C) & 535.28 & 542.77 & 547.39 
        & 535.49 & 544.14 & 549.33    
        & 533.99 & 542.23 & 543.51
        & 534.08&  542.09 & 546.99\\
++G** (S) &	535.69	&	538.56	& 539.72 
            & 536.02	& 540.8	& 542.28 
            & 534.44 & 538.61 & 539.78 
            & 534.58 &  539.06        &    540.36   \\
 ++G** (C) 
            & 535.21 & 538.08 & 539.18
            & 535.52 & 540.26 & 541.75 
            & 533.96 & 538.13 & 539.31
            & 534.08 &   538.54 &        539.83 
            \\\hline
\end{tabular}
\end{table}

\clearpage
\begin{table}[htb]
\caption{CO, Oxygen K-edge. CC2 and CCSD oscillator strengths for the first three core excitation using different standard basis sets.
\label{CO_forze_Oxygen}}
\centering
\scriptsize
\begin{tabular}{|l|ccc|ccc|} \hline
     &   \multicolumn{3}{c|}{CC2} &
         \multicolumn{3}{|c|}{CCSD} \\\cline{2-4}\cline{5-7}
Basis set & $f_1$ & $f_2$ & $f_3$ 
     & $f_1$ & $f_2$ & $f_3$ \\\hline
cc-pVDZ	&	0.0705	&	0.0016	&	0.0004	&	0.0775	&	0.0023	&	0.0017	\\
cc-pVTZ	&	0.0732	&	0	&	0.0005	&	0.0785	&	0	&	0.0023	\\
cc-pVQZ	&	0.0717	&	0.0002	&	0.0005	&	0.0782	&	0.0003	&	0.0024	\\
cc-pV5Z	&	0.0704	&	0.0008	&	0.0002	&	0.0771	&	0.0012	&	0.0008	\\\hline
aug-cc-pVDZ	&	0.0658	&	0.0008	&	0	&	0.0775	&	0.0012	&	0.001	\\
aug-cc-pVTZ	&	0.0692	&	0.0009	&	0	&	0.0777	&	0.0012	&	0.001	\\
aug-cc-pVQZ	&	0.0695	&	0.0008	&	0	&	0.0777	&	0.0012	&	0.001	\\
aug-cc-pV5Z	&	0.0696	&	0.0007	&	0	&	0.0779	&	0.0011	&	0.0011	\\\hline
d-aug-cc-pVDZ	&	0.0657	&	0.0006	&	0	&	0.077	&	0.0011	&	0.0008	\\
d-aug-cc-pVTZ	&	0.0692	&	0.0006	&	0	&	0.0775	&	0.0011	&	0.001	\\
d-aug-cc-pVQZ	&	0.0695	&	0.0006	&	0	&	0.0777	&	0.0011	&	0.001	\\
d-aug-cc-pV5Z	&	0.0696	&	0.0006	&	0	&	0.0779	&	0.0011	&	0.001	\\\hline
cc-pCVDZ	&	0.0701	&	0.0015	&	0.0004	&	0.0772	&	0.0022	&	0.0017	\\
cc-pCVTZ	&	0.073	&	0	&	0.0005	&	0.0788	&	0	&	0.0023	\\
cc-pCVQZ	&	0.0717	&	0.0002	&	0.0004	&	0.0786	&	0.0004	&	0.0023	\\
cc-pCV5Z	&	0.0704	&	0.0008	&	0.0002	&	0.0783	&	0.001	&	0.002	\\\hline
aug-cc-pCVDZ	&	0.0655	&	0.0008	&	0	&	0.0777	&	0.0012	&	0.001	\\
aug-cc-pCVTZ	&	0.0691	&	0.0009	&	0	&	0.0768	&	0.0012	&	0.0008	\\
aug-cc-pCVQZ	&	0.0696	&	0.0008	&	0	&	0.0781	&	0.0012	&	0.001	\\
aug-cc-pCV5Z	&	0.0697	&	0.0007	&	0	&	0.0782	&	0.0011	&	0.0011	\\\hline
d-aug-cc-pCVDZ	&	0.0654	&	0.0006	&	0	&	0.0767	&	0.0011	&	0.0009	\\
d-aug-cc-pCVTZ	&	0.069	&	0.0006	&	0	&	0.0777	&	0.0011	&	0.001	\\
d-aug-cc-pCVQZ	&	0.0695	&	0.0006	&	0	&	0.0781	&	0.0011	&	0.001	\\
d-aug-cc-pCV5Z	&	0.0696	&	0.0006	&	0	&	0.0782	&	0.0011	&	0.001	\\\hline
6-311G	&	0.0801	&	0.0015	&	0.0008	&	0.0841	&	0.0021	&	0.0025	\\
6-311G**	&	0.0746	&	0.0011	&	0.0011	&	0.0789	&	0.0015	&	0.0024	\\
6-311++G**	&	0.0701	&	0.0008	&	0	&	0.0777	&	0.0012	&	0.0009	\\
\hline
\end{tabular}
\end{table}

\begin{table}[ht]
\caption{NH$_3$, Nitrogen K-edge. CC2, CCSD, CC3 and CCSDT results for the first three vertical core excitations (eV). 
The experimental values (peak maxima) are $\omega_1 =400.66$ eV ($a$), 
$\omega_2 = 402.33$ eV ($e$), $\omega_3 = 402.86$~eV.~\cite{SchirmerExp}
\label{NH3_energies}}
\centering
\scriptsize
\begin{tabular}{l|ccc|ccc|ccc|ccc} \hline
     & \multicolumn{3}{c|}{CC2}
     & \multicolumn{3}{c|}{CCSD}
     & \multicolumn{3}{c|}{CC3} 
     & \multicolumn{3}{c}{CCSDT}
     \\\cline{2-4}\cline{5-7}\cline{8-10}\cline{11-13}
Basis set & $\omega_1$ & $\omega_2$ & $\omega_3$ 
          & $\omega_1$ & $\omega_2$ & $\omega_3$ 
          & $\omega_1$ & $\omega_2$ & $\omega_3$ 
          & $\omega_1$ & $\omega_2$ & $\omega_3$\\\hline
VDZ & 404.99 & 407.00 & -- 
    & 404.82 & 406.67 & 417.35 
    & 403.96 & 405.95 & 416.46
    &--&--&--\\  
VTZ & 402.30 & 404.31 & -- &401.99 & 403.85 & 409.68 & 400.88 & 402.88 & 408.56
&--&--&--\\           
VQZ & 402.04 & 403.99 & -- &401.86 & 403.68 & 408.06 &
400.55 & 402.51 & 406.71
&--&--&--\\   
V5Z & 401.83 & 403.60 & 406.38 & 401.77 & 403.53 & 406.12  & 400.36 & 402.21 & 404.88
&--&--&--   \\\hline
aVDZ & 403.41 & 404.84 & 406.43 
   & 404.10 & 405.79 & 405.79 
   & 402.83 & 404.47 & 405.98 
   & 403.02 & 404.72 & 406.16 \\ 
aVTZ & 401.44 & 402.83 & 404.03 
   & 401.62 & 403.28 & 404.39 
   & 400.26 & 401.88 & 403.05
   & 400.32 & 401.99 & 403.12\\ 
aVQZ & 401.44 & 402.82 & 403.79 
   & 401.59 & 403.24 & 404.20  
   & 400.14 & 401.75 & 402.72
   & 400.18 & 401.85 & 402.79 \\
aV5Z & 401.44 & 402.80 & 403.55 
   & 401.61 & 403.25 & 404.08
   & 400.12 & 401.73 & 402.51 
   & -- &-- &-- \\\hline
dVDZ & 403.39 & 404.71 & 405.08 & 404.08 & 405.75 & 406.39 
& 402.81 & 404.41 & 404.97
&&&\\  
dVTZ & 401.43 & 402.77 & 403.14 & 401.62 & 403.27 & 403.91 & 400.26 & 401.86 & 402.38
&&&\\
dVQZ & 401.44 & 402.78 & 403.15 & 401.59 & 403.23 & 403.88 & 400.15 & 401.74 & 402.27 &&&\\  
dV5Z & 401.44  &402.78 & 403.14 & 401.61 & 403.25 & 403.90 & 400.12 & 401.72 & 402.25 &&&\\  \hline
CVDZ & 404.49 & 406.49 & -- & 404.34 & 406.17 & 416.83 
      & 403.31 & 405.31 & 415.79 &&&\\  
CVTZ & 402.72 & 404.74 & -- & 402.50 & 404.35 & 410.18 
      & 401.31 & 403.31 & 408.99 &&&\\  
CVQZ & 402.41 & 404.36 & -- & 402.34 & 404.35 & 408.55 & 400.99 & 402.95 & 407.16 &&&\\  
CV5Z & 402.19 & 403.96 & 406.74 & 402.23  & 403.99 & 406.58 
     & 400.80 & 402.64 & 405.30 &&&\\  \hline
aCVDZ & 402.91 & 404.34 & 405.92 
    & 403.60 & 405.29 & 406.67 
    & 402.19 & 403.83 & 405.33
    & 402.33 & 404.04 &  405.47 \\
aCVTZ & 401.85 & 403.24 & 404.45 
    & 402.13 & 403.79 & 404.90 
    & 400.71 & 402.33 & 403.49
    & 400.76 & 402.44 & 403.57 \\
aCVQZ & 401.81 & 403.18 & 404.15 
    & 402.07 & 403.72 & 404.68  
    & 400.58 & 402.19 & 403.16 
    & 400.63 & 402.29 & 403.24 \\
aCV5Z & 401.78 & 403.15 & 403.90 
    & 402.06 & 403.70 & 404.53 
    & 400.54 & 402.15 & 402.94 
    & -- & -- & -- \\\hline
dCVDZ &402.89 & 404.21 & 404.57 & 403.59 &405.25 & 405.90 & 402.17 & 403.77 & 404.29\\
dCVTZ &401.85 & 403.18 & 403.55 & 402.13 &403.78 & 404.42 & 400.71 & 402.30 & 402.83\\
dCVQZ &401.80 & 403.14 & 403.50 & 402.07 &403.71 & 404.37 & 400.58 & 402.18 & 402.71\\
dCV5Z &401.78 & 403.13 & 403.48 & 402.06 &403.70 & 404.36 & -- & -- & -- \\\hline
G  & 403.08 & 405.13 & -- 
   & 402.59 & 404.52 & 410.90 
   & 401.67 & 403.75 & 410.03
   & 401.62 & 403.71 & --\\
G**  (S) & 402.94 & 404.89 & --
         & 402.48 & 404.33 & 410.56
         & 401.55 & 403.52 & 409.65
         & 401.52 & 403.50 & -- \\
G** (C) & 402.60  & 404.59 & -- 
        & 402.10  & 403.98 & 410.19 
        & 401.17  & 403.18 & 409.28
        & 401.13 & 403.14 & -- \\
++G** (S) & 401.92 & 403.42 & 405.27 
           & 402.07 & 403.76 & 405.35
           & 400.82 & 402.49 & 404.20
           & 400.90 & 402.61 & --\\
++G**(C) & 401.62 & 403.11 & 404.96   
         & 401.70 & 403.41 & 405.00
         & 400.47 & 402.15 & 403.86 
         & 400.53 &402.25& -- \\      
           \hline
\end{tabular}
\end{table}

\begin{table}[h]
\caption{NH$_3$, Nitrogen K-edge. CC2 and CCSD results for the oscillator strengths (length gauge) of the first three core excitations using different standard basis sets. \label{NH3_forze}}
\centering
\scriptsize
\begin{tabular}{|l|ccc|ccc|ccc|ccc} \hline
     & \multicolumn{3}{c|}{CC2}
     &   \multicolumn{3}{c|}{CCSD} \\\cline{2-4}\cline{5-7}
Base & $f_1$ & $f_2$ & $f_3$&  $f_1$ & $f_2$ & $f_3$ \\\hline
cc-pVDZ & 0.0069  &  0.0703&--           
  & 0.0080  &  0.0773&--   
\\
cc-pVTZ & 0.0055&0.0623&--                  
  & 0.0068&0.0718&-- \\
cc-pVQZ & 0.0047&0.0521&--      
  & 0.0062&0.0650&--\\
cc-pV5Z & 0.0045&0.0407& 0.0205  &    
  0.0063 &0.0574& 0.0226
\\\hline
aug-cc-pVDZ & 0.0032 &0.0160& 0.0129    & 
     0.0057 &0.0387& 0.0153\\
aug-cc-pVTZ & 0.0034 &0.0166& 0.0102     & 
     0.0056 &0.0395& 0.0124\\
aug-cc-pVQZ & 0.0035 &0.0162& 0.0080       & 
           0.0057  &0.0401& 0.0106\\
aug-cc-pV5Z & -- & -- & --
   & -- & -- & -- 
\\\hline
d-aug-cc-pVDZ & 0.0032  &0.0120 &0.0023      
    & 0.0057  &0.0381 &0.0062\\
d-aug-cc-pVTZ  & 0.0035  &0.0139 &0.0025      
    & 0.0057  &0.0390 &0.0064\\
d-aug-cc-pVQZ  & 0.0035  &0.0142 &0.0025       
    & 0.0057  &0.0398 &0.0064\\
d-aug-cc-pV5Z  &     --     &   --     & --
    &-- &-- &--
\\\hline
cc-pCVDZ  & 0.0068 &0.0697&--      
    & 0.0080 &0.0771&--\\
cc-pCVTZ  & 0.0055 &0.0619&--     
    & 0.0069 &0.0719&-- \\
cc-pCVQZ  & 0.0047 &0.0518&--       
    & 0.0062 &0.0654&-- \\
cc-pCV5Z  & 0.0045 &0.0404& 0.0204         
    & 0.0063 &0.0577& 0.0226\\\hline
aug-cc-pCVDZ & 0.0032 &0.0158& 0.0129        
    & 0.0057 &0.0386& 0.0153\\
aug-cc-pCVTZ & 0.0034 &0.0163& 0.01011      
    & 0.0057 &0.0397& 0.0124\\
aug-cc-pCVQZ & 0.0035 &0.0160& 0.0079      
    & 0.0058 &0.0405& 0.0106\\
aug-cc-pC5Z &        --  & --       & --&-- &-- &-- 
\\\hline
d-aug-cc-pVCDZ &0.0032 &0.0120&0.0023
    &0.0057 &0.0381&0.0063\\
d-aug-cc-pCVTZ&0.0034 &0.0136&0.0024        
    &0.0057 &0.0392&0.0064\\
d-aug-cc-pCVQZ &0.0035 &0.0140&0.0024     
    &0.0058 &0.0401&0.0064\\
d-aug-cc-pCV5Z &--     & --   & --
    & --&-- &-- 
\\\hline
6-311G &  0.0066   &0.0673&--     
  &  0.0076   &0.0761&-- \\
6-311G** & 0.0062 &0.0657  &--    
    & 0.0071 & 0.1104 &-- \\
6-311++G** (S)& 0.0036  & 0.0205 & 0.0154
      & 0.0057 & 0.0409 & 0.0167\\
6-311++G**(C) 
      &    &   &  
      & 0.0056 & 0.0408 & 0.0167\\\hline
\end{tabular}
\end{table}


\clearpage

\bibliography{Tesi}